\documentclass{article}

\usepackage{amssymb,amsfonts,amsmath}
\usepackage{cite,enumerate,float,indentfirst}
\usepackage{color}

\def\be{\begin{eqnarray}}
\def\ee{\end{eqnarray}}
\def\nn{\nonumber}

\def\p{\partial}
\def\tr{{\rm tr}\,}

\def\P{{\bf p}}

\definecolor{red}{rgb}{1,0,0}
\definecolor{orange}{rgb}{1,0.5,0}
\definecolor{violet}{rgb}{0.7,0,1}

\hoffset= - 1.0in         

\begin{document}

\hfill ITEP/TH-28/18

\hfill IITP/TH-17/18

\bigskip

\centerline{\Large{Cut-and-join operators and Macdonald polynomials from the 3-Schur
functions
}}

\bigskip

\centerline{\bf A.Morozov}

\bigskip

\centerline{\it ITEP \& IITP, Moscow, Russia}

\bigskip

\centerline{ABSTRACT}

\bigskip

{\footnotesize
Schur polynomials of infinitely many time-variables 
are among the most important special functions of 
modern mathematical physics. 
They are directly associated with the characters of
linear and symmetric group and are therefore labeled
by Young diagrams.
They possess a somewhat mysterious deformation
to Macdonald and Kerov polynomials, 
which no longer has group-theory interpretation, 
still preserves most of the nice properties of Schur functions.
The family of Schur-Macdonald function, 
however, is not big enough --
needed for various applications are counterparts of the
Schur functions, labeled by plane (3d) partitions.
Recently a very concrete suggestion was made 
on how this generalization can be done --
and miraculous coincidences on this way can serve as a support
to the idea, which, however, needs a lot of work to become
a reliable and efficient theory.
In particular,
one can expect that Macdonald and even entire Kerov deformations 
should appear in this theory on equal right with the
ordinary 2-Schur functions.
In this paper we demonstrate in some detail how 
this works for Macdonald polynomials and how they
emerge from the 3-Schur functions
when the {\it vector} time-variables, associated with plane-partitions, 
are projected onto the ordinary {\it scalar} times
under non-vanishing angles, which depend on $q$ and $t$.
We also explain how the cut-and-join operators
smoothly interpolate between different cases.
Most of consideration is restricted to level two.
}

\bigskip

\bigskip

\section{Introduction}

One of the tasks of modern theoretical physics is to make quantum field 
and string theories calculable beyond perturbation expansions.
A good understanding of non-perturbative phenomena is achieved by
development of instanton calculus in supersymmetric theories, where
perturbative effects can be almost eliminated.
Then one can look not only on the space-time dependencies, 
but on the far more interesting structures in the space of theories,
their coupling constants and boundary conditions.
These properties are best described in terms of integrable systems \cite{UFN3,GKMMM},
and efficiently handled by the formalism of matrix, tensor and network models.
However, while the simplest matrix models are already well investigated,
this is not yet the case with their generalizations.
One of the main obstacles in these studies is the need for more
general special functions with distinguished hidden-symmetry properties, 
which can be used to express the universal (common) features of many different
models and theories and can provide a language to express the
answers in quantitative form.
There are many reasons to expect that the relevant new functions should
include the generalization of $\tau$-functions in one particular direction:
from Young diagrams to plane (3d) partitions.  
Then, since for the ordinary $\tau$-functions
the starting point was the theory of Schur polynomials, 
one can ask for their generalization to 3-Schur functions, which would
depend on the properly extended set of time-variables.

To be more concrete, 
$2d$ Virasoro conformal blocks \cite{CFT} and Nekrasov functions \cite{Nekfun},
encoding the instanton effects in $4d$ SYM theories, involve sums over
Young diagrams (partitions of integers) -- and AGT equivalence \cite{AGT}
between these two types of quantities is best described by
peculiar ("conformal" or "Dotsenko-Fateev") matrix models \cite{confmamo},
efficiently handled by the theory of Schur functions.
Increasing space-time dimension leads to $q$ and $t$-deformations,
substituting Schur by Macdonald polynomials \cite{Macdonald}.
However, they are not enough for description of generic $6d$ case,
where the single-loop Virasoro is lifted to a double-loop DIM algebra \cite{DIM},
Young diagrams are substituted by plane (3d) partitions,
and matrix models are promoted to {\it network models} \cite{network},
defined on a rich variety of graphs.
Needed in this case are the new "3-Schur" functions, depending on
additional time variables --
they were recently introduced in \cite{3Schur}.\footnote{For alternative
attempt on the lines of \cite{genpols} see \cite{Z3pols}.
Perhaps, 3-Schurs are not yet the end of the story:
further generalization to $8d$ involves solid (4d) partitions,
and Nekrasov functions seem to dramatically simplify \cite{Nekmag} --
still underlying symmetries and even the relevant set of time variables,
nothing to say about the 4-Schur functions,
are not understood.}
We assume familiarity with that paper and further elaborate in one of
the many directions which is opened by discovery of this new class
of special functions.
Namely, we provide more details on embedding of the theory of
Macdonald polynomials into that of the 3-Schur functions.
We also begin developing the formalism of the cut-and-join operators
which in the case of 3-Schurs has a non-abelian extension,
not well studied even in the case of ordinary 2-Schur functions.

We now remind just a few basic things about 2-Schur functions and
the way they are lifted to the 3-Schur case.
Young diagrams are partitions of integeres, i.e. they are ordered
sequences of positive numbers, $R=[r_1\geq r_2\geq \ldots \geq r_{l_R}>0]$,
often depicted as diagrams on the plane with $l_R$ lines of boxes (squares), 
of lengths $r_1,\ldots,r_{l_R}$. 
We often call the total number of boxes $|R|=r_1+\ldots+r_{l_R}$ the {\it level}.
A generating function for all Young diagrams depends on inifnitely many
"time-variables" $p_k$ -- with diagram $R$ one associates a monomial
$p_R = \prod_{i=1}^{l_R} p_{r_i}$.
The ordinary Schur  (2-Schur) functions $S_R\{p_k\}$ are  functions of
these time-variables, which are the common eigenfunctions of the infinitely many
commuting "cut-and-join" operators $\hat W_\Delta$ \cite{MMN1}, 
also labeled by Young diagrams:
\be
\hat W_\Delta S_R\{p\} = \psi_R(\Delta) S_R\{p\}
\ee
with eigenvalues $\psi_R(\Delta)$ being the characters of symmetric group
${\cal S}_{|R|}$
(the name "cut-and-join" is inherited from the simplest of these operators
$\hat W_{[2]}$, which has a clear interpretation of this kind and appears
in a whole variety of applications).
Furthermore, the coefficients of Schur functions are again the 
symmetric-group characters: 
\be
S_R\{p\} = \sum_\Delta \frac{\psi_R(\Delta)p_\Delta}{z_\Delta}
\ee
with the standard combinatorial factors $z_\Delta$,
and orthogonality of characters imply orthogonality of Schur functions
in appropriate scalar product 
\be
\Big<p_\Delta\Big|p_{\Delta'}\Big>= z_\Delta\cdot \delta_{\Delta,\Delta'} 
\label{scaprodu}
\ee
what can be used as their alternative denition (as orthogonalization of a
monomial basis in the space of symmetric polynomials).
Schur functions form a ring with multiplication, describing that of
representations of linear group $sl_N$, and they reduce to the characters
of these representations at the Miwa locus $p_k=\tr X^k$ for $N\times N$ matrix $X$.
Among the most important properties of Schur functions is Cauchy summation
formula 
\be
\sum_R S_R\{p\} S_R\{p'\} = \exp \left(\sum_k \frac{p_kp'_k}{k}\right)
\ee
Other well known facts, like determinantal representations and Plucker/Hirota 
relations are peculiar for Schur functions and more difficult to generalize --
in neither of the three directions:
to Macdonald-Kerov \cite{MMkerov}, generalized Macdonald \cite{genpols} 
and 3-Schur \cite{3Schur} functions.

Coming straight to 3-Schur functions, they are labeled by plane partitions $\pi$,
which are the {\it two}-indexed sets of non-increasing integers,
$\pi_{i,j}\geq \pi_{i+1,j}$ and $\pi_{i,j}\geq \pi_{i,j+1}$.
They can be depicted as piles of cubes lying in the corner in 3 dimensions --
thus they are sometime called 3-partitions.
The generating functions of such partitions require  
an extended set of time variables
$\vec p_k = p_k^{(a)}$ with additional index $a=1,\ldots,k$,
i.e. dimension of the vector $\vec p_k$ is equal to its "level" (grading) $k$.
The 3-Schur are functions of these time-variables, and this means 
that   ${\cal S}_\pi\{\vec p_k\}$
at level $m=|\pi|$ are described by the rank-$m$ tensors,
what relates this theory to another rapidly developing new branch of
theoretical physics -- the theory of {\it tensor models} \cite{tenmod}.
Both stories are particular parts of the underestimated, still successful
{\it non-linear algebra} program \cite{NLA},
targeted at extending {\it all} the results of the linear algebra from matrices
to arbitrary tensors.

In \cite{3Schur} the 3-Schur functions were defined by orthogonalization
w.r.t. the direct analogue of the scalar product (\ref{scaprodu}),
for vector-valued $\vec p$-variables.
Despite these 3-Schur functions do not have an immediate relation to group-theory
(though we definitely expect them to play a role in description of DIM representations),
they {\it are} the common eigenfunctions of an extended set of cut-and-join operators.
Moreover, these operators are easily reduced not only to the standard one
for Schur functions, but also to a deformation, which provides the
Ruijsenaars Hamiltonians for Macdonald polynomials -- therefore
the technique of $\hat W$-operators is convenient for the study of
3-Schurs reduction to Macdonald functions.
Developing this technique will be one of the main purposes of the present paper.
In fact,  commuting operators form a kind of Cartanian part of a much larger
algebra of $\hat W$-operators, which acts between 3-Schur functions at each level.
One of the interesting open questions is their unification into quantities,
naturally acting on all levels at once - what in the case of Schur and Macdonald
functions goes all the way down to the Sugawara/Cherednik-Dunkl construction.

We will also continue studying the other properties of 3-Schur functions,
in particular, provide an evidence that they satisfy the Cauchy summation formula.
This adds to the mysteries of 3-Schur functions 
(like resolvability of overdefined system of orthogonality relations or linear 
dependencies between emerging $\vec p$-vectors, 
which are crucial for the entire program to work, see \cite{3Schur} and the
present paper).
Like in \cite{3Schur}, we use the simplest examples at level $|\pi|=2$
to reveal the emerging phenomena
and then demonstrate that they survive in transition
to at least the next level 3.
Extension to higher levels will be worked out elsewhere.

\section{On the choice of time-variables for Macdonald polynomials}

We begin with a simple general comment about Macdonald polynomials
\cite{Macdonald}.
As usual in modern theory and its applications \cite{UFN3},
we consider them not just as symmetric
polynomials of the Miwa variables $x_a$, but as functions of
infinitely-many time-variables $p_k$, which reduce to $\{x_a\}$
on the Miwa locus $p_k = \tr X^k = \sum_a x_a^k$.
Level of sophistication depends on the choice of time-variables,
and for different purposes different choices are convenient.
We will need and use a rather unconventional one, which we now describe.

After a simple rescaling of time variables
\be
p_k \ \longrightarrow \ \frac{(1-t)^k}{1-t^k}\cdot p_k, \nn \\
M_{[n]} \ \longrightarrow \
\left(\frac{1-q}{1-t}\right)^n \prod_{i=1}^{n}\frac{(1-q^{i-1}t)}{1-q^i}\cdot M_{[n]}
\ee
symmetric and antisymmetric Macdonald polynomials acquire a simple form:
\be
M_{[n]}= {\rm Schur}_{[n]} \left\{\frac{(1-q)^k}{1-q^k}p_k\right\} \nn \\
M_{[1^n]} =  {\rm Schur}_{[1^n]} \left\{\frac{(1-t^{-1})^k}{1-t^{-k}}p_k\right\}
\ee
Other polynomials, however, remain more sophisticated.

In what follows we will apply {\it another} rescaling,
which does not change time-variables at $t=q$, i.e. for ordinary Schur functions,
but makes their scalar product
and Cauchy formula independent of $q$ and $t$:
\be
\left<p_k|p_l\right> \ = \  k\delta_{ik}
\label{metricforMac}
\ee
Then $\left<M_R|M_{R'}\right>\ \sim \ \delta_{R,R'}$
and
\be
\sum_R \frac{M_R\{p\}M_R\{p'\}}{\left<M_R|M_R\right>} =
\exp\left(\sum_k \frac{p_kp_k'}{k}\right)
\label{CauchyMac}
\ee
This makes particular expressions more complicated,
see  (\ref{Maclevel2}) for the first example.
However, this is the most convenient choice to study relation to
3-Schur functions.

\section{Cut-and-join operators for  2-Schur and Macdonald functions}

Another piece of the standard theory which we need
is that of the cut-and-join operators \cite{MMN1}.
The name comes from the simplest example,
\be
\hat W_{[2]} = \sum_{k,l} \left((k+l) p_kp_l\frac{\p}{\p p_{k+l}}+
klp_{k+l}\frac{\p^2}{\p p_k\p p_l}\right)
= \frac{1}{2}:\tr (X\p_X)^2:
\ee
which already shows up in different branches of science.
In fact such operators exist for arbitrary Young diagram
$\Delta = [\ldots, 3^{m_3}, 2^{m_2}, 1^{m_1}]$:
\be
\hat W_\Delta \sim \ :\,\tr \prod_i (X\p_X)^{m_i}\,:\
\sim \ \prod_i \hat W_{[m_i]} + \ldots
\ee
though its expression through the time-variables is more involved.
Combinatorial factor, mentioned in the Introduction, is
$z_\Delta = \prod_a a^{m_a} \cdot m_a!$

One of the many properties of these operators is that they all commute
and thus have the common eigenfunctions -- which are exactly the
Schur functions:
\be
\hat W_\Delta S_R = \psi_R(\Delta) S_R
\ee
with eigenvalues, which are the characters $\psi_R(\Delta)$
of the universal symmetric group ${\bf S}_\infty$
(available through the command {\it Chi} in MAPLE\{combinat\}
-- unfortunately this is no so for the operators themselves).
At particular level $|R|$ only the first $|R|$ time-variables,
i.e. $p_k$ with $k\leq |R|$, are operative, and operators
$\hat W_\Delta$ are accordingly and self-consistently reduced.
Moreover, only $|\Delta|\leq |R|$ are relevant -- other operators
reduce to zero.
In addition to this sub-ring of commuting  and hermitian operators
$\hat W_\Delta$  (which we naturally call {\it Cartanian}),
there are operators which convert one Schur function into another
and form a non-abelian structure --
not yet attracting attention, which it deserves.

In the simplest case of level two, which will be at the center
of our consideration in this paper, there are just two Schur functions
\be
S_{[2]} = \frac{p_2+p_1^2}{2} = S_{[2]}^+ \ \ \ \ {\rm and} \ \ \ \
S_{[1,1]} = \frac{-p_2+p_1^2}{2} = S_{[2]}^-
\label{2Schurlevel2}
\ee
where at the r.h.s. we introduced a notation, which will be consistent
with lifting to plane partitions.
These two polynomials are the common eigenfunctions of dilatation operator
and cut-and-join operators:
\be
\hat W_{[1]} = \sum_k p_k\p_k = p_1\p_1 + 2p_2\p_2 + \ldots,
&\hat W_{[1]}S_{[2]} = 2S_{[2]}, \ \ \hat W_{[1]}S_{[1,1]} = 2S_{[1,1]}, \ \ \nn \\
\hat W_{[2]} = \frac{p_2}{2}\p_1^2 + p_1^2\p_2 + \ldots,
&\hat W_{[2]}S_{[2]} = S_{[2]}, \ \ \hat W_{[2]}S_{[1,1]} = -S_{[1,1]},
\label{odW2}
\ee
\vspace{-0.3cm}
\be
\hat W_{[1,1]} = \frac{1}{2}\hat W_{[1]}(\hat W_{[1]}-1) =
p_2\p_2 + \frac{1}{2}p_1^2\p_1^2 + 2p_2p_1\p_2\p_1 + 2p_2^2\p_2^2 + \ldots,
\ \ \
&\hat W_{[1,1]}S_{[2]} = S_{[2]}, \ \  \hat W_{[1,1]}S_{[1,1]} = S_{[1,1]}
\nn\ee
where dots denote terms with $p_{k\geq 3}$, unobservable at the level 2.
The raising and lowering operators at this level are
\be
\hat W^+_{2} = \frac{p_2+p_1^2}{4}\cdot(\p_1^2 -2\p_2)  +\ldots,
 \ \ \ \ \ \ \ \ \ \ \ \ \ \ \
\hat W^+_{2}S_{[2]} = 0, \ \ \ \ \hat W^+_{2}S_{[1,1]} = S_{[2]}
\ee
and its hermitian conjugate
\be
\hat W^+_{2} = \frac{-p_2+p_1^2}{4}\cdot( \p_1^2 +2\p_2)  +\ldots,
 \ \ \ \ \ \ \ \ \ \ \ \ \ \ \
\hat W^-_{2}S_{[2]} = S_{[1,1]}, \ \ \ \  \hat W^-_{2}S_{[1,1]} = 0
\ee
In this particular example these operators look obvious,
but this is not the case in general -- and especially after the
deformations, which will be our main concern.

For Macdonald polynomials one often considers {\it difference} operators,
but in fact the {\it differential} cut-and-join operators also exist -- and these
are more relevant for our consideration at this stage.
After one more rescaling of $p_2$ in the level-two polynomials
\be
M_{[2]} \sim M_2^+=
\frac{1}{2}\left(\sqrt{\frac{(1-q)(1+t )}{(1+q)(1-t )}}\cdot p_2+p_1^2\right),
\ \ \ \ \
M_{[11]} = M_2^- =
\frac{1}{2}\left(-\sqrt{\frac{(1+q)(1-t )}{(1-q)(1+t )}}\cdot p_2 + p_1^2\right)
\label{Maclevel2}
\ee
the simplest of these operators acquires the form
\be
\widehat{WM}_{[2]}=\frac{p_2}{2}\p_1^2 + p_1^2\p_2 - \underline{2\sigma p_2\p_2} + \ldots,
 \ \ \ \ \sigma =  \frac{q-t }{\sqrt{(1-q^2)(1-t^2)}}
 = \frac{\sqrt{q/t}-\sqrt{t/q}}{\sqrt{(q-q^{-1})(t-t^{-1})}}
\label{WM2level2}
\ee
Its two eigenvalues in the space of the level-2 operators are
\be
\sqrt{\frac{(1-q)(1+t )}{(1+q)(1-t )}}
= i\sqrt{\frac{(1-q)(1+t^{-1} )}{(1+q)(1-t^{-1} )}}
\ \ \ \ \ {\rm and} \ \ \ \ \
-\sqrt{\frac{(1+q)(1-t )}{(1-q)(1+t )}}
=i\sqrt{\frac{(1+q)(1-t^{-1} )}{(1-q)(1+t^{-1} )}}
\ee

Not surprisingly, the theory of cut-and-join operators can be lifted
to the 3-Schur level.
Like 3-Schur functions themselves, they will be enlarged to a set of
commuting (Cartanian) operators, labeled by plane partitions --
and 3-Schurs will be their common eigenfunctions.
Non-abelian extension also survives and the corresponding non-hermitian
operators convert 3-Schur functions into other 3-Schurs.
Moreover,
now there is no distinguished ordering between partitions of a given size
(dualism is substituted by trialism),
"highest and lowest weights" like $S_{[r]}$ and $S_{[1^r]}$ get more abundant,
and the number of such operators
increase, together with increase of the Cartanian sub-ring.

\section{Differential cut-and-join operators. Level $2$}

\subsection{3-Schur functions}

As already mentioned, we assume familiarity with ref.\cite{3Schur}
and do not repeat the arguments and definitions from that paper.
To simplify the formulas we will use $p_2,\tilde p_2$ instead of $p_2^{(1)}$ and
$p_2^{(2)}$.

At level 2 there are three plane partitions and three 3-Schur functions
\be
\boxed{
{\cal S}_{[2]}^0 = \frac{\sqrt{2}\tilde p_2+p_1^2}{2}, \ \ \ \
{\cal S}_{[2]}^{\pm} = \frac{\pm\sqrt{\frac{3}{2}}p_2-\frac{1}{\sqrt{2}}\tilde p_2+p_1^2}{2}
}
\label{3Schurlevel2}
\ee
They are the eigenfunctions of the two linear-independent
(and, more symmetrically, three -- but linear dependent) cut-and-join operators
\be
\hat {\cal W}_{[2]}^0 = -\hat {\cal W}_{[2]}^{0'} -\hat {\cal W}_{[2]}^{0''}
=  \frac{  p_2}{2}\p_1^2 + p_1^2 \p_2 -
\underline{
\frac{1}{\sqrt{2}}(\tilde p_2\p_2+p_2\tilde\p_2)
}
+ \ldots
\label{calW20}
\ee
and
\be
\hat {\cal W}_{[2]}^{0\bot}
= \frac{1}{\sqrt{3}}\Big(\hat {\cal W}_{[2]}^{0''} -\hat {\cal W}_{[2]}^{0'}\Big) =
\frac{\tilde  p_2}{2}\p_1^2 + p_1^2\tilde\p_2 - \frac{1}{\sqrt{2}}(p_2\p_2-\tilde
p_2\tilde\p_2)
+ \ldots
\label{W20bot}
\ee
The dots, to be omitted below, stand for the terms with $p_n$, $n\geq 3$,
which are irrelevant at level $2$.
Superscript $0$ means that these are "Cartanian" generators, which leave the 3-Schur
functions intact:
\be
\hat{\cal W}_{[2]}^0 {\cal S}_{[2]}^0 = 0,  \ \ \ \
\hat{\cal W}_{[2]}^0 {\cal S}_{[2]}^\pm = \pm \sqrt{\frac{3}{2}} {\cal S}_{[2]}^\pm,
\nn\\
\hat{\cal W}_{[2]}^{0,\bot} {\cal S}_{[2]}^0 = \sqrt{2},  \ \ \ \
\hat{\cal W}_{[2]}^{0,\bot} {\cal S}_{[2]}^\pm = - \frac{1}{\sqrt{2}} {\cal S}_{[2]}^\pm
\ee

There are also additional pairs/triples of $W$ with superscripts $\pm$,
which act as raising and lowering operators, for example
\be
\hat{\cal W}_{[2]}^\pm =
p_1^2\Big(-\frac{1}{2}\frac{\p}{\p  p_2}
\mp \frac{\sqrt{3}}{2}\frac{\p}{\p  \tilde p_2}\Big)
+ \frac{-  p_2\pm\sqrt{3}\tilde p_2}{4} \frac{\p^2}{\p p_1^2}
- \frac{1}{\sqrt{2}}\Big(\tilde p_2\frac{\p}{\p p_2} + p_2\frac{\p}{\p \tilde p_2}\Big)
+ \ldots
\ee
act as raising and lowering generators in the chain $(+0-)$:
\be
\hat{\cal W}_{[2]}^+ {\cal S}_{[2]}^+ = 0,  \ \ \ \
\hat{\cal W}_{[2]}^+ {\cal S}_{[2]}^0 = -\sqrt{\frac{3}{2}}{\cal S}_{[2]}^+,
\ \ \ \ \hat{\cal W}_{[2]}^+ {\cal S}_{[2]}^- = \sqrt{\frac{3}{2}}{\cal S}_{[2]}^0
\nn \\
\hat{\cal W}_{[2]}^- {\cal S}_{[2]}^+ = -\sqrt{\frac{3}{2}}{\cal S}_{[2]}^0,  \ \ \ \
\hat{\cal W}_{[2]}^- {\cal S}_{[2]}^0 = \sqrt{\frac{3}{2}}{\cal S}_{[2]}^-,
\ \ \ \ \hat{\cal W}_{[2]}^+ {\cal S}_{[2]}^- = 0
\ee
There are two more pairs of operators -- for the chains $(0-+)$ and $(-+0)$
(i.e. the corresponding raising operators nullify respectively
${\cal S}_{[2]}^0$ and ${\cal S}_{[2]}^-$).
They can be obtained by $\pm\frac{2\pi}{3}$ rotations in the plane $(p_2,\tilde p_2)$.
Thus, together with the triple of Cartanian $\hat {\cal W}_{[2]}^0$
acting at level 2 is the non-abelian set of $1+9$ operators (not all linear independent),
$1$ stands for the dilatation
\be
\hat {\cal W}_{[1]}= \sum_k k\vec p_k\vec\p_k =\sum_k \sum_{a=1}^k
kp_k^{(a)}\frac{\p}{\p p_k^{(a)}}
\ee
which was observable already at level one.

\subsection{Relation/reduction of 3-Schur to ordinary 2-Schur functions}

Interpolation from $\hat {\cal W}_{[2]}^0$ in (\ref{calW20})
to the ordinary (2-Schur) operator (\ref{odW2}),
\be
\hat W_{[2]}=\frac{  p_2}{2}\p_1^2 + p_1^2 \p_2
\ee
is straightforward:
\be
\hat W_{[2]}^0(h) = \frac{p_2}{2}\p_1^2 + p_1^2\p_2 -
{h(\tilde p_2\p_2+p_2\tilde\p_2)}
\ee
The three interpolating eigenfunctions  are
\be
{\cal S}^\lambda_{[2]}(h)\sim
(1-\lambda^2 )\tilde p_2+h\lambda  p_2 + hp_1^2
\label{interpolSchurlevel2}
\ee
with eigenvalues $\lambda$ which are the three roots of
the characteristic equation
\be
\lambda(\lambda^2-1-h^2)=0
\ee
i.e. the three functions are
\be
{\cal S}^{ 0}_{[2]}(h) \sim hp_1^2+\tilde p_2, \ \ \
{\cal S}^{ \pm }_{[2]}(h) \sim -h\tilde p_2 \pm \sqrt{1+h^2}\,p_2+p_1^2
\ee
and in the 2-Schur limit the first of them depends only on the "foreign"
time-variable $\tilde p_2$.

The $h$-deformation of the second operator (\ref{W20bot}) is
\be
\frac{1}{\sqrt{2}}\hat W_{[2]}^{0,\bot}(h) =
h\Big(\frac{\tilde p_2}{2}\p_1^2 + p_1^2\tilde\p_2\Big)
- h^2p_2\p_2 + (1-h^2)\tilde p_2\tilde\p_2
\ee
at $h\longrightarrow 0$ turns into $\tilde p_2\tilde\p_2$, i.e. acts only
on the "hidden" coordinate $\tilde p_2$ and "disappears" (decouples)
from the world of the 2-Schur functions:
it has eigenvalues $-h^2$ for ${\cal S}_{[2]}^{\pm}(h)$, which survive in this limit,
and $1$ for the decoupling ${\cal S}^0_{[2]}(h)$.

Note that all these level-two cut-and-join operators
consist of three independent combinations
$\frac{p_2}{2}\p_1^2+p_1^2\p_2$,\  $p_2\tilde\p_2+\tilde p_2\p_2$ and
$p_2\p_2-\tilde p_2\tilde\p_2$, which are hermitian (self-conjugate)
in the 3-Schur generalization
\be
<p_k^{(a)}|p_l^{(b)}>\ =\ k\delta_{k,l}
\label{metricfor3Schur}
\ee
of the metric (\ref{metricforMac}).
In this metric ${\vec p_k}\!^\dagger = k\vec \p_k$ and
${\vec\p_k}\!^\dagger = \frac{1}{k}\vec p_k$.
The forth hermitian combination $p_2\p_2+\tilde p_2\tilde\p_2$
is a part of the dilatation operator
\be
\hat W^0_{[1]} = \sum_k k\vec p_k\vec\p_k
\ee
which acts already at the level one.

\subsection{Rotation in the $p_2$-plane}

Clearly, the structures of the last (underlined) terms
in $\widehat{WM}_{[2]}$ in (\ref{WM2level2})
and $\hat {\cal W}_{[2]}^0$ in (\ref{calW20}) are different,
so there can be no direct interpolation between {\it them}.
The rescue comes from existence of two other cut-and-join operators
for 3d-Schurs,
\be
\hat{{\cal W}_2^{0}}' = p_1^2\Big(-\frac{1}{2}\frac{\p}{\p p_2} -
\frac{\sqrt{3}}{2}\frac{\p}{\p \tilde p_2}\Big)
- \frac{ p_2+\sqrt{3}\tilde  p_2}{4} \frac{\p^2}{\p p_1^2}
+ \frac{1}{2\sqrt{2}}\left(
(  \sqrt{3}    p_2+\tilde p_2)\frac{\p}{\p   p_2}
+ (p_2-\sqrt{3}  \tilde p_2 ) \frac{\p}{\p\tilde  p_2}
\right) \nn \\
\hat{{\cal W}_2^{0}}'' = p_1^2\Big(-\frac{1}{2}\frac{\p}{\p p_2} +
\frac{\sqrt{3}}{2}\frac{\p}{\p \tilde p_2}\Big)
- \frac{ p_2-\sqrt{3}\tilde  p_2}{4} \frac{\p^2}{\p p_1^2}
+ \frac{1}{2\sqrt{2}}\left(
(  -\sqrt{3}    p_2+\tilde p_2)\frac{\p}{\p   p_2}
+ (p_2+\sqrt{3}  \tilde p_2 ) \frac{\p}{\p\tilde  p_2}
\right)
\ee
of which we can take a linear combination:
\be
- u\cdot \hat {{\cal W}_{[2]}^0}' - v\cdot \hat {{\cal W}_{[2]}^0}'' =
u\cdot \left\{
\frac{p_2+\sqrt{3}\tilde p_2 }{4}
\left(\frac{ -\sqrt{3} \p_2+\tilde \p_2}{\sqrt{2}}+\p_1^2\right)
+ \left(\frac{-\sqrt{3}  p_2+\tilde p_2}{2\sqrt{2}}+p_1^2\right)
\frac{\p_2+\sqrt{3}\tilde\p_2}{2}
\right\} + \nn \\
+v\cdot \left\{
\frac{p_2-\sqrt{3}\tilde p_2}{4}
\left(\frac{\sqrt{3} \p_2+\tilde \p_2}{\sqrt{2}}+\p_1^2\right)
+ \left(\frac{ \sqrt{3} p_2+\tilde p_2}{2\sqrt{2}}+p_1^2\right)
\frac{\p_2-\sqrt{3} \tilde \p_2}{2}
\right\} = \nn \\
= \frac{(cp_2+s\tilde p_2)}{2}\p_1^2 + p_1^2(c  \p_2+s\tilde \p_2)
-\frac{1}{\sqrt{2}}\Big(c(\tilde p_2\p_2+p_2\tilde \p_2)
+ s(p_2\p_2 -\tilde p_2\tilde\p_2)\Big)
\ee
Here we introduced the rotation parameters are $c=\frac{u+v}{2}=\cos\theta$
and $s=\frac{\sqrt{3}(u-v)}{2}=\sin\theta$.
For $c=1$ and $s=0$ we return back to $\hat{\cal W}_{[2]}^0$.
The 3-Schur functions are still the eigenfunctions of this operator,
but the eigenvalues are now
\be
\lambda^0=s\sqrt{2} \ \ \  {\rm for} \ \  {\cal S}^0_{[2]}
\ \ \ \ \ {\rm and} \ \ \ \ \
\lambda^{\pm} =\frac{-s\pm c\sqrt{3}}{\sqrt{2}} \ \ \  {\rm for} \ \ {\cal S}^\pm_{[2]}
\ee
We can now switch to the new coordinates in the space $(p_2,\tilde p_2)$:
\be
 P_2 = cp_2+s\tilde p_2, \ \ \   \tilde P_2 =-s p_2+c\tilde p_2 \nn \\
  D_2 = c  \p_2+s\tilde \p_2, \ \ \  \tilde  D_2 =-s  \p_2+c\tilde \p_2
\ee
and obtain:
\be
\hat{\cal W}^\theta_{[2]} :=
\frac{  P_2}{2} \p_1^2 + p_1^2  D_2
- \frac{\cos(3\theta)}{\sqrt{2}}(\tilde P_2 D_2+P_2\tilde D_2)
-\frac{\sin(3\theta)}{\sqrt{2}}(P_2D_2-\tilde P_2\tilde D_2 )
\ee
where $\cos(3\theta) =  (c^2-3s^2)c=(1-4s^2)c$ and $\sin(3\theta) =  (3c^2-s^2)s =
(3-4s^2)c$.
Orthogonal operator is
\be
\hat{\cal W}^{\theta\bot}_{[2]}=
 \frac{\tilde P_2}{2} \p_1^2 + p_1^2  \tilde D_2
+\frac{ \sin(3\theta) }{\sqrt{2}} (\tilde P_2 D_2+P_2\tilde D_2)
-\frac{\cos(3\theta)}{\sqrt{2}}  (P_2D_2 - \tilde P_2\tilde D_2)
\label{calWbot}
\ee
and eigenvalues are equal to $\lambda_\bot
=\sqrt{2}\cdot\frac{ 1-\frac{\sin(3\theta)}{\sqrt{2}}\lambda-\lambda^2}{\cos(3\theta)}$:
\be
\lambda^0=\sqrt{2} \sin\theta, \ \ \ \ \ \ \
&\lambda^\pm = \sqrt{2}\sin\left(\theta\pm\frac{2\pi}{3}\right)
\nn\\
 \lambda_\bot^0   = \sqrt{2}\cos\theta,
\ \ \ \ \ \
& \lambda_\bot^\pm
=  \sqrt{2}\cos\left(\theta\pm\frac{2\pi}{3}\right)
\label{evrot}
\ee
Note that two eigenvalues $\lambda$ coincide when, say, $\theta=\frac{\pi}{6}$,
but the corresponding $\lambda_\bot$ do not:
denominator in the mapping $\lambda \longrightarrow \lambda_\bot$
vanishes at such points, and l'Hopital's resolution provides different $\lambda_\bot$,
according to (\ref{evrot}).

Another remark is that the mapping $\theta\longrightarrow 3\theta$ is exactly the one,
which solves generic cubic equation:
\be
x^3-bx-c=0 \ \Longrightarrow \ x =
\sqrt{\frac{4b}{3}}\sin\left(\theta+\frac{2k\pi}{3}\right),
\ k=0,\pm 1
\ \ \ \ \ {\rm provided} \ \ \sqrt{\frac{27c^2}{4b^3}} = \sin(3\theta)
\ee
Thus it is not surprising that eigenvalues are nicely described
at level $n=2$, when the number of 3-Schur functions is three.
However, at higher levels $n>2$, the number of functions and
the order of characteristic equation for the eigenvalues $\lambda$ increase,
then the existence of trigonometric solutions requires it to be of a special type.

\subsection{Interpolation/reduction from 3-Schur to MacDonald polynomials}

Now we possess the structure, which appears in the last term in (\ref{WM2level2}),
and can interpolate to Macdonald polynomials.
It is enough to multiply $\tilde P_2$ and $\tilde D_2$
by $h\sqrt{2}$ and adjust the rotation angle
so that $\sin(3\theta)= 2\sqrt{2}\sigma$, i.e.
\be
\boxed{
\sin(3\theta)
= 2\sqrt{2}\sigma
=  2\sqrt{2}\frac{(q-t)}{\sqrt{(1-q^2)(1-t^2)}}
=  2\sqrt{2}i\,\frac{1-qt^{-1}}{\sqrt{(1-q^2)(1-t^{-2})}}
 =  2\sqrt{2}\frac{\sqrt{q/t}-\sqrt{t/q}}{\sqrt{(q-q^{-1})(t-t^{-1})}}
 }
\label{thetathroughsigma}
\ee

In other words
\be
\boxed{
\tilde{\cal W}^\theta_{[2]}(h)=
\frac{ P_2}{2} \p_1^2 + p_1^2  D_2
- h\cos(3\theta) (\tilde P_2 D_2+P_2\tilde D_2)
-\frac{\sin(3\theta)}{\sqrt{2}}(P_2D_2 - 2h^2\tilde P_2\tilde D_2 )
}
\label{Wthetahlevel2}
\ee
interpolates between
\be
\tilde{\cal W}^\theta_{[2]}
= \frac{ P_2}{2} \p_1^2 + p_1^2  D_2
- \frac{\cos(3\theta)}{\sqrt{2}} (\tilde P_2 D_2+P_2\tilde D_2)
-\frac{\sin(3\theta)}{\sqrt{2}}(P_2D_2-\tilde P_2\tilde D_2 )
\ee
at $h=\frac{1}{\sqrt{2}}$
and
\be
\widehat{WM}_{[2]}=\frac{  P_2}{2}\p_1^2 + p_1^2  D_2
-\frac{\sin(3\theta)}{\sqrt{2}}  P_2  D_2
\ee
at $h=0$.
Their eigenfunctions
\be
\boxed{
{\cal S}^\lambda_{[2]}(\theta,h) =
h\cos(3\theta)\Big(\lambda   P_2 + p_1^2\Big)
+ \left(1-\frac{\sin(3\theta)}{\sqrt{2}}\lambda-\lambda^2\right) \tilde P_2
}
= h\cos(3\theta)\Big(\lambda   P_2 + p_1^2\Big)  - \frac{\lambda h^2\cos^2(3\theta)}
{\lambda-h^2\sqrt{2}\sin(3\theta)}\tilde P_2 =
\nn
\ee
\vspace{-0.4cm}
\be
= h\cos(3\theta)p_1^2 + \left(\lambda h\cos(3\theta)\cos\theta +
\frac{\lambda h^2\cos^2(3\theta)\sin\theta}
{\lambda-h^2\sqrt{2}\sin(3\theta)}\right)p_2
+ \left(h\lambda\cos(3\theta)\sin\theta
- \frac{\lambda h^2\cos^2(3\theta)\cos\theta}
{\lambda-h^2\sqrt{2}\sin(3\theta)}  \right)\tilde p_2
\label{Slamhthelevel2}
\ee
are the $\theta$-deformations of   (\ref{interpolSchurlevel2}),
and the eigenvalues $\lambda$ are now defined as the three roots of
the $(h,\theta)$-deformed characteristic equation

\be
\ \ \ \ \ \ \ \ \ \ \ \ \ \ \ \ \
\boxed{
\Big(\lambda-h^2\sqrt{2}\sin(3\theta)\Big)
\left(\lambda^2+\frac{\sin(3\theta)}{\sqrt{2}}\lambda - 1\right)
= \lambda h^2\cos^2(3\theta)
}
\label{cubelambda}
\ee
{\footnotesize
\be
\begin{array}{ccccc}
&&\!\!\!\!\!\!\!\!\!\!\!\!\!\!\! \!\!\!\!\!\!\!\!\!\!\!\!\!\!\! \uparrow\\  \\
&&
\!\!\!\!\!\!\!\!\!\!\!\!\!\!\!\!\!\!\!\!\!\!\!\!\!\!\!\!\!\!\!\!\!\!\!\!\!\!\!\!\!\!\!\!\!\!\!\!
\sqrt{2}\lambda(\lambda^2-1-h^2) + \sin(3\theta)\Big(\lambda^2(1-2h^2)+2h^2\Big) = 0
\!\!\!\!\!\!\!\!\!\!\!  && \\
\\
&\!\!\!\!\!\!\!\!\!\!\!\!\!\!\!\!\!\!\!\!\!\!\!\!\!\!\!\!
\!\!\!\!\!\!\!\!\!\!\!\!\!\!\! \!\!\!\!\!\!\!\!\!\!\!\!\!\!\!
h=\frac{1}{\sqrt{2}}\ \ \ {\rm (3-Schurs)}\ \swarrow
&&
\!\!\!\!\!\!\!\!\!\!\!\!\!\!\!\!\!\!\!\!\!\!\!\!\!\!\!\!\!\!\!\!\!\!\!\!
\searrow \ h=0 \ \ \ {\rm (2-Schurs)}\\ \\
\sqrt{2}\lambda(2\lambda^2-3)=-2\sin(3\theta)
=2 s(4s^2-3)
\!\!\!\!\!\!\!\!\!\!\!\!\!\!\!\!\!\!\!\!\!\!\!\!\!\!\!\!\!\!\!\!\!\!
&&&
\!\!\!\!\!\!\!\!\!\!\!
\sqrt{2}\lambda\Big(\lambda^2  +\frac{\sin(3\theta)}{\sqrt{2}}\lambda-1\Big)
= \sqrt{2}\lambda(\lambda-\lambda_+)(\lambda-\lambda_-)=0
\\ \\
\lambda = \sqrt{2}s=\sqrt{2}\sin\theta,
\ \ \ -\frac{1}{\sqrt{2}}s \pm\sqrt{\frac{3}{2}}c
= \sqrt{2}\sin\Big(\theta\pm\frac{2\pi}{3}\Big)
\!\!\!\!\!\!\!\!\!\!\!\!\!\!\!\!\!\!\!\!\!\!\!\!\!\!\!\!\!\!\!\!\!\!\!\!\!\!
\!\!\!\!\!\!\!\!\!\!\!\!\!\!\!\!\!\!\!\!\!\!
&&&
\!\!\!\!\!\!\!\!
 \lambda=0, \ \ \
\lambda_+=\sqrt{\frac{(1-q)(1+t)}{(1+q)(1-t)}}, \ \ \
\lambda_-=-\sqrt{\frac{(1+q)(1-t)}{(1-q)(1+t)}}
\end{array}
\nn
\ee
}

\noindent
Eigenfunctions (\ref{Slamhthelevel2})
are $\theta$-deformations of (\ref{interpolSchurlevel2}) and
interpolate between the triple of 3-Schur functions  at $h=\frac{1}{\sqrt{2}}$
\be
\frac{1}{2h\cos(3\theta)}{\cal S}^{\lambda=\sqrt{2}\sin\theta}_{[2]}
\left(\theta,h=\frac{1}{\sqrt{2}}\right)
= \frac{1}{\sqrt{2}}(\sin\theta P_2+\cos\theta \tilde P_2) + \frac{1}{2}p_1^2
=\frac{\sqrt{2}\tilde p_2+p_1^2}{2} ={\cal S}_{[2]}^0,
\ee
\vspace{-0.3cm}
\be
\frac{1}{2h\cos(3\theta)}{\cal S}^{\lambda=
-\frac{1}{\sqrt{2}}\sin\theta\pm\sqrt{\frac{3}{2}}\cos\theta}_{[2]}
\left(\theta,h=\frac{1}{\sqrt{2}}\right)
= \frac{-s\pm c\sqrt{3}}{2\sqrt{2}}P_2
-\frac{c\pm s\sqrt{3}}{2\sqrt{2}}\tilde P_2 + \frac{1}{2}p_1^2
P_2
=\pm\sqrt{\frac{3}{2}}  p_2-\frac{1}{\sqrt{2}}\tilde p_2+p_1^2={\cal S}_{[2]}^\pm
\nn
\ee
and the pair of MacDonald polynomials  (\ref{Maclevel2})
at $h=0$:
\be
\lim_{h\longrightarrow 0}
\frac{1}{2h\cos(3\theta)}{\cal S}^{\lambda_\pm}_{[2]}
\left(\theta,h=0\right)
= \frac{1}{2}\left(\lambda_\pm P_2 + p_1^2\right)=M_{[2]}^\pm
\ee
In the latter case the third function becomes independent of $P_2$ and $p_1$
\be
{\cal S}^{\lambda=0}_{[2]}\left(\theta,h=0\right) =   \tilde P_2
\ee
and "decouples".

\subsection{More on deformed cut-and-join operators}

A function of the form (\ref{Slamhthelevel2}) with arbitrary (unrelated)
$\lambda,h$ and $\theta$
is annihilated by three operators:
\be
\frac{P_2}{2}\p_1^2 + p_1^2P_2   + (\lambda-\lambda^{-1})P_2D_2 + \lambda \tilde P_2\tilde
D_2
- \lambda, \nn \\
\frac{\tilde P_2}{2}\p_1^2 + p_1^2\tilde P_2   + \mu\lambda P_2D_2
+ \Big(\mu\lambda-(\mu\lambda)^{-1}\Big) \tilde P_2\tilde D_2 - \mu\lambda
\ee
and
\be
P_2\tilde D_2+\tilde P_2 D_2 - \mu P_2D_2-\mu^{-1}\tilde P_2\tilde D_2
\ee
with $\mu =
\frac{1-\frac{\sin(3\theta)}{\sqrt{2}}\lambda-\lambda^2}{h\lambda\cos(3\theta)}$.
However, we want an operator which annihilates three such functions with
three different $\lambda$ at once -- or, to be more precise, have them as eigenfunctions.
This means that we can add, say the third operator with some $\lambda$-independent
coefficient $w$ to the first one, and all the terms, except for the $p$-independent one,
should not depend on $\lambda$.
There are two such terms, $uP_2D_2+v P_2\tilde D_2$, with
\be
u = \lambda-\frac{1}{\lambda} - w \mu, \ \ \ \ v = \lambda - \frac{w}{\mu}
\ \ \ \ \Longrightarrow \ \ \ \
(\lambda-v)(\lambda^2-u\lambda-1)=\lambda w^2
\ee
Comparing this with the cubic equation (\ref{cubelambda}) for $\lambda$, we conclude that
$u = -\frac{\sin(3\theta)}{\sqrt{2}}, \ \ \ v= \sqrt{2}h^2\sin(3\theta), \ \ \
w= -h\cos(3\theta)$,
i.e. reproduce the operator (\ref{Wthetahlevel2}):
\be
\boxed{
\tilde{\cal W}^\theta_{[2]}(h)=
\frac{ P_2}{2} \p_1^2 + p_1^2  D_2
- h\cos(3\theta) (\tilde P_2 D_2+P_2\tilde D_2)
-\frac{\sin(3\theta)}{\sqrt{2}}(P_2D_2 - 2h^2\tilde P_2\tilde D_2 )
}
\label{Wthetahlevel2a}
\ee
Similarly, adding the third operator to the second one, we get:
\be
u^\bot = \mu\lambda - w^\bot \mu, \ \ \ \ v^\bot = \mu\lambda -\frac{1}{\mu\lambda}
- \frac{w^\bot}{\mu}
\ \ \ \ \Longrightarrow \ \ \ \
(\mu\lambda-u^\bot)(\mu^2\lambda^2-v^\bot\mu\lambda-1)=\mu\lambda (w^\bot)^2
\ee
Thus we can expect that $\mu\lambda$ also satisfies a cubic equation -- what is indeed
the case, because its inverse
$\frac{1}{\mu\lambda} \ \stackrel{(\ref{cubelambda})}{=}\
-\frac{\lambda-h^2\sqrt{2}\sin(3\theta)}{\lambda h\cos(3\theta)}\ $
is a linear function of $\lambda^{-1}$, which satisfies a cubic equation.
It remains to write it down explicitly:
\be
\mu = \frac{1-\frac{\sin(3\theta)}{\sqrt{2}}\lambda-\lambda^2}{h\lambda\cos(3\theta)}
\ \stackrel{(\ref{cubelambda})}{=} \
-\frac{h\cos(3\theta)}{\lambda-h^2\sqrt{2}\cos(3\theta)} \ \ \
\Longrightarrow \nn \\
\Longrightarrow \ \ \
\Big(\mu\lambda +h\cos(3\theta\Big)
\left(\mu^2\lambda^2  - \frac{1-h^2-2h^4\sin^2(3\theta)}{h\cos(3\theta)}\cdot \mu\lambda
-1\right)
=  2h^4\sin^2(3\theta) \mu\lambda
\ee
what implies that
$u^\bot = -h\cos(3\theta), \ \ \
v^\bot= \frac{1-h^2-2h^4\sin^2(3\theta)}{h\cos(3\theta)}, \ \ \
w^\bot =\sqrt{2}h^2\sin(3\theta)$,
and the second operator, supplementing (\ref{Wthetahlevel2}), is
{\footnotesize
\be
\boxed{
\tilde{\cal W}^{\theta\bot}_{[2]}(h)\sim
h\cos(3\theta)\left(\frac{\tilde P_2}{2} \p_1^2 + p_1^2  \tilde D_2\right)
+\frac{h^3\sin(6\theta)}{\sqrt{2}} (\tilde P_2 D_2+P_2\tilde D_2)
-h^2\cos^2(3\theta)  P_2D_2 +\Big(1-h^2-2h^4\sin^2(3\theta)\Big)\tilde P_2\tilde D_2
}
\label{Wbotthetahlevel2}
\nn
\ee
}
As usual, it gets independent of $P_2$ and $p_1$ in the limit $h=0$,
and is unseen at the level of Macdonald polynomials.

The eigenvalues of
this operator are
\be
\lambda_\bot = h\cos(3\theta)\cdot \mu\lambda
= 1-\frac{\sin(3\theta)}{\sqrt{2}}\lambda-\lambda^2
\label{lambot}
\ee
and solve the cubic equation:
\be
\lambda_\bot^3 - (1-2h^2)\Big(1+h^2\sin^2(3\theta)\Big)\lambda_\bot^2
+ h^2(h^2-2)\cos^2(3\theta)\lambda_\bot - h^4\cos^4(3\theta)=0
\label{lamboteigenvalues}
\ee
which, after substitution of (\ref{lambot}), is, of course, equivalent to
(\ref{cubelambda}).

At symmetric (i.e. the 3-Schur) point $h=\frac{1}{\sqrt{2}}$ the entire operator
is divisible by $\cos(3\theta)$
(thus in (\ref{W20bot}) and (\ref{calWbot}) this factor was omitted),
and the eigenvalues are given by rescaled (\ref{evrot}):
\be
\begin{array}{ll}
\frac{\lambda^0}{\sqrt{2}} = \sin\theta, \ \ \ \ \ \ \
&\frac{\lambda^\pm }{\sqrt{2}} =\sin\left(\theta\pm\frac{2\pi}{3}\right)
\\ \\
\frac{\lambda_{\bot_{}}^0}{\cos(3\theta)} =
\frac{(1-\sin^2\theta)(1-4\sin^2\theta)}{\cos(3\theta)}
= \cos\theta,
\ \ \ \ \ \
&\frac{\lambda_{\bot_{}}^\pm}{\cos(3\theta)}
=  -\frac{(1-4\sin^2\theta)\cdot\cos\theta\cdot\frac{\cos\theta\pm \sqrt{3}\sin\theta}{2}}
{\cos(3\theta)}
=  \cos\left(\theta\pm\frac{2\pi}{3}\right)
\end{array}
\ee

\subsection{$h$-evolution in pictures}

To visualize the situation, we can use the following picture:

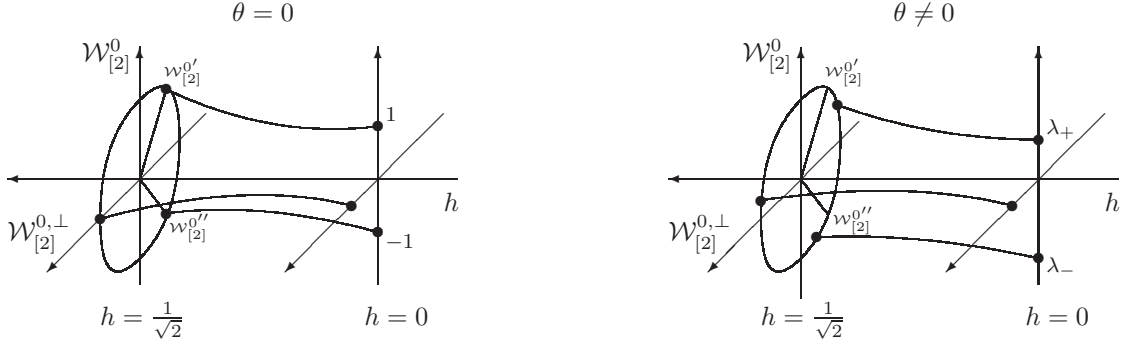
\begin{picture}(400,140)(-250,-60)

\put(-200,0){
\put(120,0){\vector(-1,0){170}}    \put(115,-12){\mbox{$h$}}
\put(0,-40){\vector(0,1){90}}     \put(-22,45){\mbox{${\cal W}^0_{[2]}$}}
\put(25,25){\vector(-1,-1){60}}   \put(-50,-23){\mbox{${\cal W}^{0,\bot}_{[2]}$}}

\qbezier(0,-30)(-15,-45)(-15,-15)
\qbezier(0,30)(-15,15)(-15,-15)
\qbezier(0,-30)(15,-15)(15,15)
\qbezier(0,30)(15,45)(15,15)

\put(-15,-15){\circle*{4}}
\put(10,34){\circle*{4}}
\put(10,-13){\circle*{4}}
\qbezier(0,0)(5,17)(10,34)       \put(9,39){\mbox{{\tiny ${\cal W}^{0'}_{[2]}$}}}
\qbezier(0,0)(5,-6.5)(10,-13)    \put(11,-20){\mbox{{\tiny ${\cal W}^{0''}_{[2]}$}}}

\put(90,-40){\vector(0,1){90}}
\put(115,25){\vector(-1,-1){60}}
\put(80,-10){\circle*{4}}
\put(90,20){\circle*{4}}     \put(93,22){\footnotesize\mbox{$1$}}
\put(90,-20){\circle*{4}}    \put(93,-26){\footnotesize\mbox{$-1$}}

\put(85,-55){\mbox{$h=0$}}
\put(-15,-55){\mbox{$h=\frac{1}{\sqrt{2}}$}}

\put(35,60){\mbox{$\theta=0$}}

\qbezier(-15,-15)(40,-0)(80,-10)
\qbezier(10,34)(50,15)(90,20)
\qbezier(10,-13)(45,-8)(90,-20)

}

\put(50,0){
\put(120,0){\vector(-1,0){170}}    \put(115,-12){\mbox{$h$}}
\put(0,-45){\vector(0,1){95}}     \put(-22,45){\mbox{${\cal W}^0_{[2]}$}}
\put(22,22){\vector(-1,-1){57}}   \put(-50,-23){\mbox{${\cal W}^{0,\bot}_{[2]}$}}

\qbezier(0,-30)(-15,-45)(-15,-15)
\qbezier(0,30)(-15,15)(-15,-15)
\qbezier(0,-30)(15,-15)(15,15)
\qbezier(0,30)(15,45)(15,15)

\put(-15,-8){\circle*{4}}
\put(14,28){\circle*{4}}
\put(6,-22){\circle*{4}}
\qbezier(0,0)(5,17)(10,34)   \put(9,39){\mbox{{\tiny ${\cal W}^{0'}_{[2]}$}}}
\qbezier(0,0)(5,-6.5)(10,-13)    \put(12,-17){\mbox{{\tiny ${\cal W}^{0''}_{[2]}$}}}

\put(90,-40){\vector(0,1){90}}
\put(115,25){\vector(-1,-1){60}}
\put(80,-10){\circle*{4}}
\put(90,15){\circle*{4}}     \put(93,17){\footnotesize\mbox{$\lambda_+$}}
\put(90,-30){\circle*{4}}    \put(93,-35){\footnotesize\mbox{$\lambda_-$}}

\put(85,-55){\mbox{$h=0$}}
\put(-15,-55){\mbox{$h=\frac{1}{\sqrt{2}}$}}

\put(35,60){\mbox{$\theta\neq0$}}

\qbezier(-15,-8)(40,-0)(80,-10)
\qbezier(14,28)(50,15)(90,15)
\qbezier(6,-22)(45,-20)(90,-30)
}

\end{picture}

\noindent
Black dots denote triples of eigenvalues of the two Cartanian operators
at different values of deformation parameters $h$ (horizontal axis) and $\theta$
(two pictures).
At $h=0$ on the vertical axis we get eigenvalues
for Schur (at $\theta=0$) and Macdonald (at generic $\theta$) functions.
At $h=\frac{1}{\sqrt{2}}$ these are the eigenvalues of the 3-Schur functions
in original coordinates $\vec p_2$ and $\theta$-rotated $\vec P_2$.
Rotation is actually along a circle in the $(p_2,\tilde p_2)$ plane,
which can be associated with a circle in the vertical plane of $\hat W$ eigenvalues.
Similarly, at $h=0$ the $p_2$ axis, relevant for description of
ordinary 2-Schur and Macdonald functions can be associated with a vertical line.

Interpolation, which we described above, connects the black dots at fully-$3d$
$Z_3$-symmetric point $h=\frac{1}{\sqrt{2}}$ with those at the $2d$-point $h=0$
in a straightforward way.
However, while this can seem natural at $\theta=0$, when $\theta$ increases towards
$\frac{2\pi}{3}$, it is natural to expect that twisting will result in
a jump to another branch of interpolating function --
what is not yet immediately  seen in our formulas.

Some information is provided by discriminant of (\ref{cubelambda})
\be
{\rm disc}_\lambda\left\{\lambda(\lambda^2-1-h^2)
+ \frac{\sin(3\theta)}{\sqrt{2}}\Big(\lambda^2(1-2h^2)+2h^2\Big)\right\}
= \nn\\
= 4(1+h^2 )^3 + \frac{\sin^2(3\theta)}{2}(1-38h^2 -75h^4 +76h^6 + 4h^8)
-2h^2(1-2h^2)^3\sin^4(3\theta)
\label{discW0}
\ee
but it is limited, because vanishing of this discriminant means only that
the projections on the vertical line of the "trajectories"  $\lambda^\theta(h)$
intersect -- and this is necessary, but not sufficient for a true intersection.
For example, as obvious from the picture, for $\sin(3\theta)=1$,
say, $\theta = \frac{\pi}{6}$,
this discriminant vanishes already at the symmetric (3d) point
$h=\frac{1}{\sqrt{2}}$:
\be
\left.{\rm disc}_\lambda(\ref{cubelambda})\right|_{\sin(3\theta)=1}
\sim (1+h^2)^2(1-2h^2)^2
\ee
Indeed, vertical projections obviously coincide at this point,
but the points themselves remain different --
coincident are the two eigenvalues of ${\cal W}^{0}_{[2]}$,
but not of ${\cal W}^{0,\bot}_{[2]}$, and no reshuffling takes place.

Technically, it is more practical to plot $h^2$ as a function of $\lambda$:
and then rotate the picture:
\be
 \lambda(\lambda^2-1-h^2) + \frac{\sin(3\theta)}{\sqrt{2}}\Big(\lambda^2(1-2h^2)+2h^2\Big) =
0
\ \ \ \Longrightarrow \ \ \
h^2 = \frac{\lambda(\lambda-\lambda_+)(\lambda-\lambda_-)}
{\lambda+\sqrt{2}(\lambda^2-1)\sin(3\theta) }
\ee
where $\lambda_- = -\lambda_+^{-1}$ and \ $\sin(3\theta)=-\sqrt{2}(\lambda_+ +\lambda_-)$.
Then we easily get the following pattern from the evolution of $\lambda$ with $h^2$
for $\frac{1}{\sqrt{2}}<|\lambda_+|<\sqrt{2}$, i.e. $ |\sin(3\theta)|<1$:

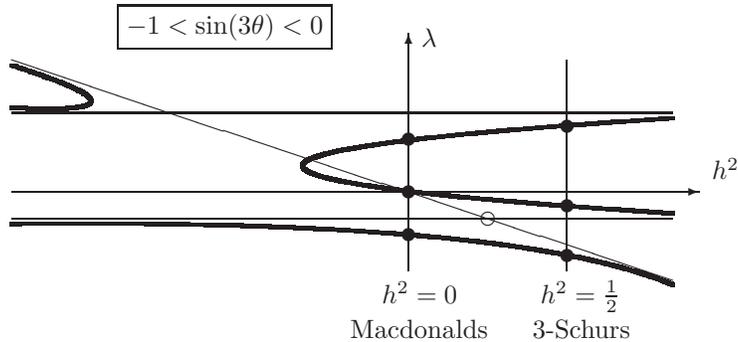
\begin{picture}(300,150)(-200,-70)

\put(-110,60){\mbox{$\boxed{-1<\sin(3\theta)<0}$}}

\put(-150,0){\vector(1,0){260}}
\put(-150,-10){\line(1,0){250}}
\put(-150,30){\line(1,0){250}}

\put(0,-30){\vector(0,1){90}}
\put(60,-30){\line(0,1){80}}
\put(-150,50){\line(3,-1){250}}

\put(0,0){\circle*{5}}\put(0,20){\circle*{5}}\put(0,-16){\circle*{5}}
\put(60,-5){\circle*{5}}\put(60,25){\circle*{5}}\put(60,-24){\circle*{5}}

\put(30,-10){\circle{5}}

\put(115,5){\mbox{$h^2$}}
\put(-10,-42){\mbox{$h^2=0$}} \put(-22,-55){\mbox{\rm Macdonalds}}
\put(50,-42){\mbox{$h^2=\frac{1}{2}$}} \put(47,-55){\mbox{\rm 3-Schurs}}

\put(5,55){\mbox{$\lambda$}}

\linethickness{0.5mm}
\qbezier(-150,-12)(40,-10)(100,-35)
\qbezier(-150,48)(-90,29)(-150,32)
\qbezier(100,28)(-180,10)(100,-8)

\end{picture}

\noindent
This particular picture is for positive $1<\rho<\sqrt{2}$,
i.e. for negative $\sin(3\theta)$,
the switch to positive $\sin(3\theta)$ via
the changes $\rho\longrightarrow \rho^{-1}$
or $\rho\longrightarrow -\rho$ is upside-down reflection:

\noindent
The region $h^2<0$ is "unphysical", but it is there that
the behaviour of evolution lines is defined.
This particular picture is for positive $1<\lambda_+<\sqrt{2}$,
i.e. for negative $\sin(3\theta)$,
the switch to positive $\sin(3\theta)$ via
the changes $\lambda_+\longrightarrow \lambda_+^{-1}=-\lambda_-$
or $\lambda_+\longrightarrow -\lambda_+=\lambda_-^{-1}$ is upside-down reflection:

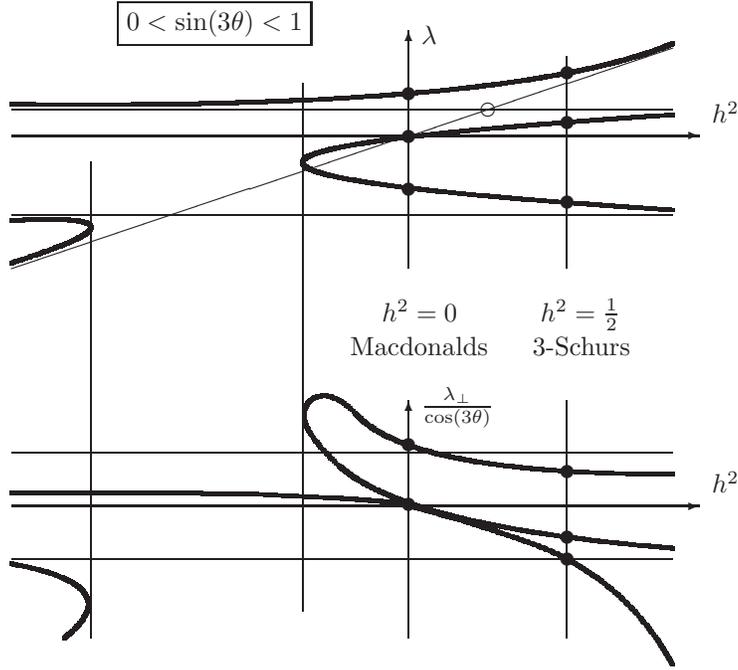
\begin{picture}(300,270)(-200,-210)

\put(-110,40){\mbox{$\boxed{0<\sin(3\theta)<1}$}}

\put(-150,0){\vector(1,0){260}}
\put(-150,10){\line(1,0){250}}
\put(-150,-30){\line(1,0){250}}

\put(0,-50){\vector(0,1){90}}
\put(60,30){\line(0,-1){80}}
\put(-150,-50){\line(3,1){250}}

\put(0,0){\circle*{5}}\put(0,-20){\circle*{5}}\put(0,16){\circle*{5}}
\put(60,5){\circle*{5}}\put(60,-25){\circle*{5}}\put(60,24){\circle*{5}}

\put(30,10){\circle{5}}

\put(115,5){\mbox{$h^2$}}
\put(-10,-70){\mbox{$h^2=0$}} \put(-22,-83){\mbox{\rm Macdonalds}}
\put(50,-70){\mbox{$h^2=\frac{1}{2}$}} \put(47,-83){\mbox{\rm 3-Schurs}}

\put(5,35){\mbox{$\lambda$}}


\put(0,-140){

\put(-150,0){\vector(1,0){260}}
\put(-150,20){\line(1,0){250}}
\put(-150,-20){\line(1,0){250}}

\put(0,-50){\vector(0,1){90}}
\put(60,40){\line(0,-1){90}}
\put(-120,-50){\line(0,1){180}}
\put(-40,-40){\line(0,1){200}}

\put(0,0.5){\circle*{5}} \put(0,23){\circle*{5}}
\put(60,13){\circle*{5}}\put(60,-20){\circle*{5}}\put(60,-12){\circle*{5}}


\put(115,5){\mbox{$h^2$}}

\put(5,35){\mbox{$\frac{\lambda_{\bot_{}}}{\cos(3\theta)}$}}

\linethickness{0.5mm}
\qbezier(-150,5)(-60,5)(-50,4)\qbezier(-50,4)(-2,2)(0,0)
\qbezier(0,0)(20,-5)(40,-12)\qbezier(40,-12)(80,-24)(100,-60)
\qbezier(100,-16)(0,-8)(-26,15)\qbezier(-26,15)(-45,32)(-37,40)
\qbezier(-37,40)(-30,45)(-20,35)\qbezier(-20,35)(0,12)(100,12)
\qbezier(-150,-22)(-105,-30)(-130,-50)

}

\linethickness{0.5mm}
\qbezier(-150,12)(40,10)(100,35)
\qbezier(-150,-48)(-90,-29)(-150,-32)
\qbezier(100,-28)(-180,-10)(100,8)

\end{picture}

\noindent
We added also a plot for the $h$-evolution of $\frac{\lambda_{\bot_{}}}{\cos(3\theta)}$
from (\ref{lamboteigenvalues}), which does not depend on the sign of $\theta$.
Division by $\cos(3\theta)$ makes the picture more informative, by resolving the
zero of $\lambda_{\bot}$ at $\sin(3\theta)=\pm 1$, which is common for all branches.

The slope of the asymptotic line is $\sqrt{2}\sin(3\theta)$,
at vanishing $\theta$ this line becomes horizontal axis $\lambda=0$.
The two horizontal asymptotic lines are at the roots of
$\lambda+\sqrt{2}(\lambda^2-1)\sin(3\theta)$, i.e. at
\be
\lambda^\pm_{\rm as} = -\frac{1\pm \sqrt{1+8\sin^2(3\theta)}}{2\sqrt{2}\sin(3\theta)}
\ee
For vanishing $\theta$, i.e. when $t=q$,
these turn into zero and infinity, while the eigenvalues at $h=0$
become $0,\pm 1$, associated with the ordinary 2-Schur functions (\ref{2Schurlevel2}),
and at $h=\frac{1}{\sqrt{2}}$ they become $0,\pm\sqrt{\frac{3}{2}}$,
associated with original (non-rotated) 3-Schurs (\ref{3Schurlevel2}).
Note in passing, that nicely expressed through $q$ and $t$ is
a surprisingly similar, still different combination
\be
\sqrt{1+\frac{\sin^2(3\theta)}{8} } \ \stackrel{(\ref{thetathroughsigma})}{=} \
\frac{1-qt}{\sqrt{(1-q^2)(1-t^2)}}
\ee

The turning points are the roots of the equation $\frac{\p h^2}{\p \lambda} = 0$,
i.e. of $\left(\lambda^4-\frac{3}{2}\lambda^2 + 1\right)\sin(3\theta)+
\sqrt{2}\lambda\left(\lambda^2-\sin^2(3\theta)\right)$.

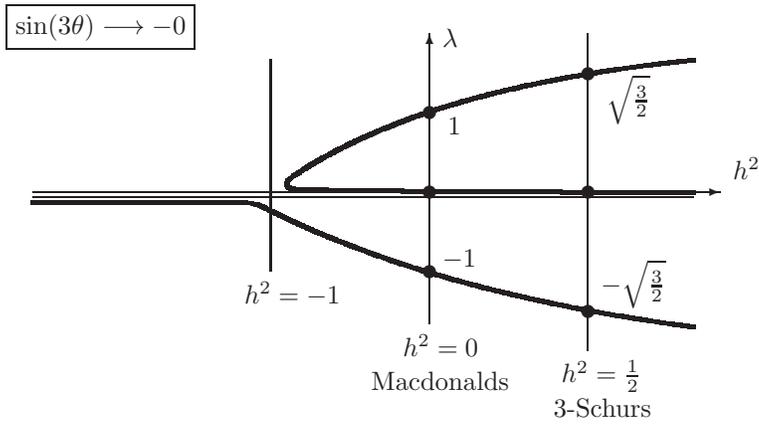
\begin{picture}(300,180)(-200,-100)

\put(-160,60){\mbox{$\boxed{\sin(3\theta)\longrightarrow -0}$}}

\put(-150,0){\vector(1,0){260}}
\put(-150,-2){\line(1,0){250}}

\put(0,-50){\vector(0,1){110}}
\put(60,-60){\line(0,1){120}}
\put(-60,-30){\line(0,1){80}}

\put(0,0){\circle*{5}}\put(0,30){\circle*{5}}\put(0,-30){\circle*{5}}
\put(60,0){\circle*{5}}\put(60,45){\circle*{5}}\put(60,-45){\circle*{5}}


\put(115,5){\mbox{$h^2$}}
\put(-10,-62){\mbox{$h^2=0$}} \put(-22,-75){\mbox{\rm Macdonalds}}
\put(50,-72){\mbox{$h^2=\frac{1}{2}$}} \put(47,-85){\mbox{\rm 3-Schurs}}
\put(-70,-42){\mbox{$h^2=-1$}}

\put(5,55){\mbox{$\lambda$}}

\put(7,22){\mbox{$1$}}  \put(5,-28){\mbox{$-1$}}
\put(67,32){\mbox{$\sqrt{\frac{3}{2}}$}}  \put(65,-38){\mbox{$-\sqrt{\frac{3}{2}}$}}

\linethickness{0.5mm}
\qbezier(-150,-4)(-80,-4)(-70,-4)\qbezier(-70,-4)(-67,-4)(-64,-5)\qbezier(-64,-5)(0,-38)(100,-51)
\qbezier(100,0)(0,0)(-50,1) \qbezier(-50,1)(-58,2)(-50,7)\qbezier(-50,7)(0,40)(100,50)

\end{picture}

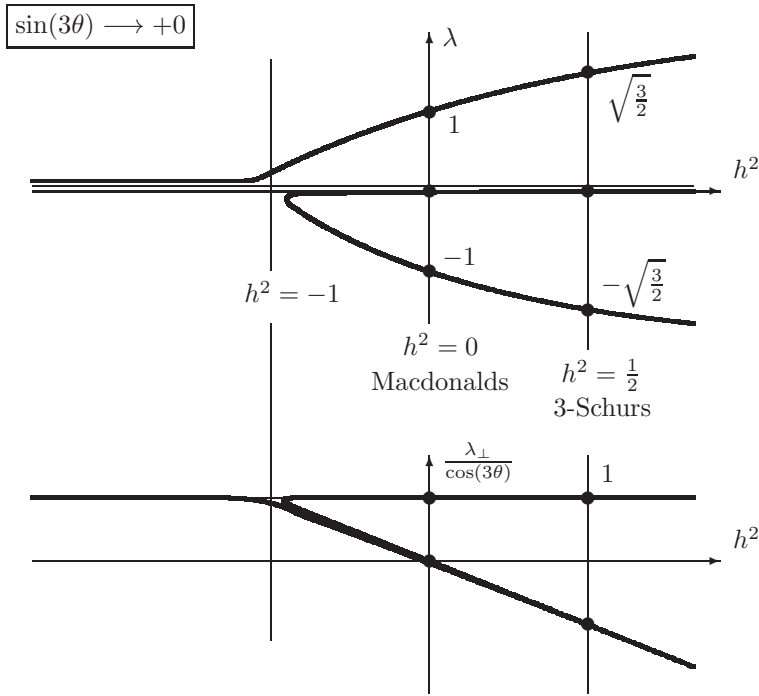
\begin{picture}(300,280)(-200,-200)

\put(-160,60){\mbox{$\boxed{\sin(3\theta)\longrightarrow +0}$}}

\put(-150,0){\vector(1,0){260}}
\put(-150,2){\line(1,0){250}}

\put(0,-50){\vector(0,1){110}}
\put(60,-60){\line(0,1){120}}
\put(-60,-30){\line(0,1){80}}

\put(0,0){\circle*{5}}\put(0,30){\circle*{5}}\put(0,-30){\circle*{5}}
\put(60,0){\circle*{5}}\put(60,45){\circle*{5}}\put(60,-45){\circle*{5}}


\put(115,5){\mbox{$h^2$}}
\put(-10,-62){\mbox{$h^2=0$}} \put(-22,-75){\mbox{\rm Macdonalds}}
\put(50,-72){\mbox{$h^2=\frac{1}{2}$}} \put(47,-85){\mbox{\rm 3-Schurs}}
\put(-70,-42){\mbox{$h^2=-1$}}

\put(5,55){\mbox{$\lambda$}}

\put(7,22){\mbox{$1$}}  \put(5,-28){\mbox{$-1$}}
\put(67,32){\mbox{$\sqrt{\frac{3}{2}}$}}  \put(65,-38){\mbox{$-\sqrt{\frac{3}{2}}$}}

\put(0,-140){

\put(-150,0){\vector(1,0){260}}
\put(-150,24){\line(1,0){250}}

\put(0,-50){\vector(0,1){90}}
\put(60,40){\line(0,-1){90}}
\put(-60,-30){\line(0,1){120}}

\put(0,0){\circle*{5}} \put(0,24){\circle*{5}}
\put(60,24){\circle*{5}}\put(60,-24){\circle*{5}}


\put(115,5){\mbox{$h^2$}}

\put(5,35){\mbox{$\frac{\lambda_{\bot_{}}}{\cos(3\theta)}$}}

\put(65,30){\mbox{$1$}}

\linethickness{0.5mm}
\qbezier(-150,24)(-80,24)(-80,24)\qbezier(-80,24)(-60,24)(-50,18)
\qbezier(-50,18)(0,0)(100,-40)
\qbezier(100,24)(0,24)(-50,24)\qbezier(-50,24)(-60,24)(-50,20)
\qbezier(-50,20)(0,0)(100,-40)

}

\linethickness{0.5mm}
\qbezier(-150,4)(-80,4)(-70,4)\qbezier(-70,4)(-67,4)(-64,5)\qbezier(-64,5)(0,38)(100,51)
\qbezier(100,0)(0,0)(-50,-1)
\qbezier(-50,-1)(-58,-2)(-50,-7)\qbezier(-50,-7)(0,-40)(100,-50)

\end{picture}

When $\sin(3\theta)$ approaches $-1$, the curve tends to the asymptotes,
moreover, the crossing, marked by a white circle, tends to $h^2=\frac{1}{2}$
(since $\lambda_{as}^- \longrightarrow \frac{1-3}{2\sqrt{2}}=-\sqrt{2}\cdot \frac{1}{2}$)
so that the two negative eigenvalues $\lambda$'s at the 3-Schur point $h^2=\frac{1}{2}$
coincide: these are $\sqrt{2}\sin\left(-\frac{\pi}{6}\right)=-\frac{1}{\sqrt{2}}$
and $\sqrt{2}\sin(\left(-\frac{\pi}{6}-\frac{2\pi}{3}\right)
= \sqrt{2}\sin\left(-\frac{5\pi}{6}\right)=-\frac{1}{\sqrt{2}}$.
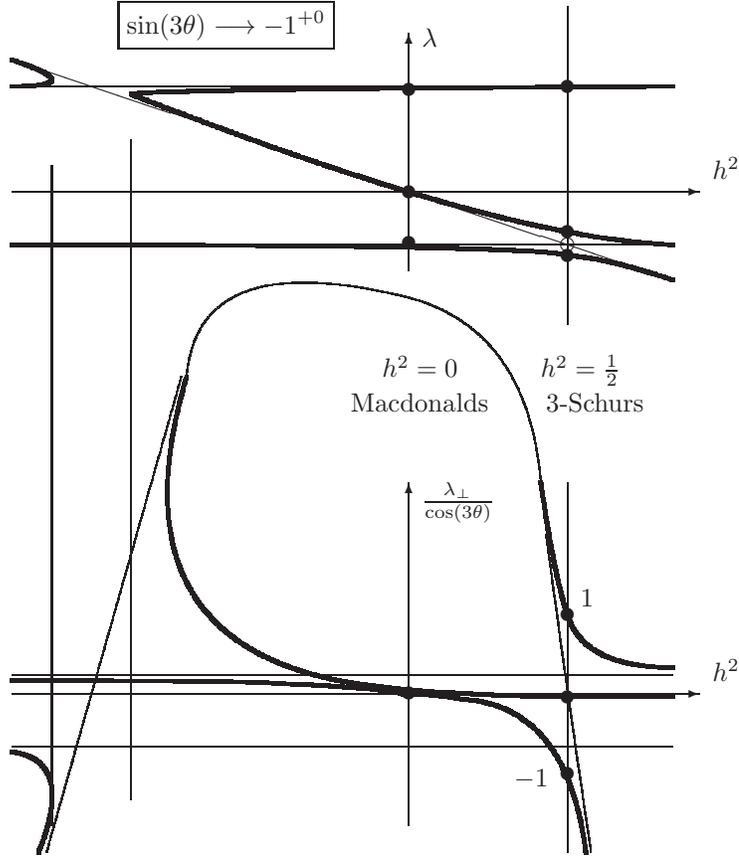
\begin{picture}(300,370)(-200,-270)

\put(-110,60){\mbox{$\boxed{ \sin(3\theta)\longrightarrow -1^{+0}}$}}

\put(-150,0){\vector(1,0){260}}
\put(-150,-20){\line(1,0){250}}
\put(-150,40){\line(1,0){250}}

\put(0,-30){\vector(0,1){90}}
\put(60,-50){\line(0,1){120}}
\put(-150,50){\line(3,-1){250}}

\put(0,0){\circle*{5}}\put(0,39){\circle*{5}}\put(0,-19){\circle*{5}}
\put(60,-15){\circle*{5}}\put(60,40){\circle*{5}}\put(60,-24){\circle*{5}}

\put(60,-20){\circle{5}}

\put(115,5){\mbox{$h^2$}}
\put(-10,-70){\mbox{$h^2=0$}} \put(-22,-83){\mbox{\rm Macdonalds}}
\put(50,-70){\mbox{$h^2=\frac{1}{2}$}} \put(52,-83){\mbox{\rm 3-Schurs}}

\put(5,55){\mbox{$\lambda$}}


\put(0,-190){

\put(-150,0){\vector(1,0){260}}
\put(-150,7){\line(1,0){250}}
\put(-150,-20){\line(1,0){250}}

\put(0,-50){\vector(0,1){130}}
\put(60,80){\line(0,-1){140}}
\put(-135,-50){\line(0,1){250}}
\put(-105,-40){\line(0,1){250}}

\put(0,0.5){\circle*{5}}
\put(60,30){\circle*{5}}\put(60,-30){\circle*{5}}\put(60,-1){\circle*{5}}


\put(115,5){\mbox{$h^2$}}

\put(5,70){\mbox{$\frac{\lambda_{\bot_{}}}{\cos(3\theta)}$}}

\qbezier(69,-60)(60,0)(52,60)
\qbezier(-137,-60)(-120,0)(-86,120)

\qbezier(-84,120)(-80,170)(0,150)\qbezier(0,150)(45,138)(50,80)
\put(65,33){\mbox{$1$}}  \put(40,-35){\mbox{$-1$}}

\linethickness{0.5mm}
\qbezier(-150,5)(-60,5)(-50,4)\qbezier(-50,4)(-2,2)(0,-1)
\qbezier(0,0)(10,0)(20,-2)\qbezier(20,-2)(60,-4)(67,-60)
\qbezier(100,-1)(80,-1)(55,-1)\qbezier(55,-1)(0,0)(-26,5)\qbezier(-26,5)(-113,22)(-84,120)
\qbezier(50,80)(55,45)(60,30)\qbezier(60,30)(66,10)(100,10)
\qbezier(-150,-22)(-126,-25)(-140,-60)

}

\linethickness{0.5mm}
\qbezier(-150,-20)(50,-20)(60,-24)\qbezier(60,-24)(75,-25)(100,-33.3)
\qbezier(-150,50)(-120,40)(-150,40)
\qbezier(100,40)(-100,39)(-105,37)\qbezier(-105,37)(0,-4)(60,-15)\qbezier(60,-15)(80,-19)(100,-20)

\end{picture}

\noindent
In the last picture for $\lambda_\bot$
the turning point between the nearly vertical branches is actually
much higher than schematically shown, it tends to infinity when $\sin(3\theta)=1$.
Also, as already mentioned, at this point all the branches of the true $\lambda_{\bot}$
actually vanish at $h^2=\frac{1}{2}$ --
we eliminated this semi-artificial zero by division over $\cos(3\theta)$.
Still inclusion of $\cos(3\theta)$ seems important for smooth interpolation,
and this singularity at the codimension-two point
$\left(\sin(3\theta)=1, \ h^2=\frac{1}{2}\right)$
in the moduli space of 3-Schur deformations does not look easily unavoidable.

Nothing special happens at $\sin(3\theta)=1$ at the Macdonald point $h=0$, still
further increase of $\lambda_+$ beyond $\sqrt{2}$ requires analytical continuation to
$|\sin(3\theta)|>1$, when eigenvalues at $h^2=\frac{1}{2}$ are no longer real
(alternatively one can consider complex-valued eigenvalues at $h=0$, i.e.
complex-valued $q$ and $t$).

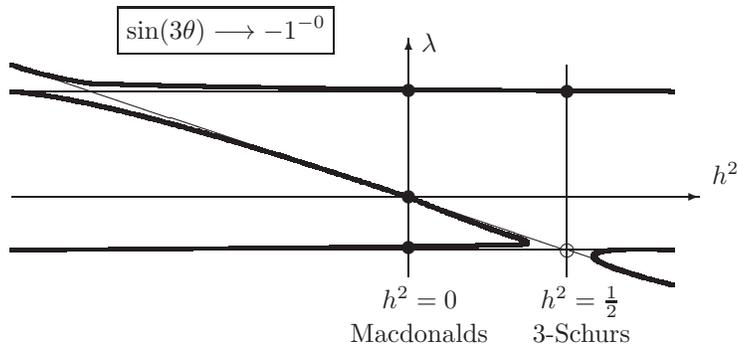
\begin{picture}(300,140)(-200,-60)

\put(-110,60){\mbox{$\boxed{ \sin(3\theta)\longrightarrow -1^{-0}}$}}

\put(-150,0){\vector(1,0){260}}
\put(-150,-20){\line(1,0){250}}
\put(-150,40){\line(1,0){250}}

\put(0,-30){\vector(0,1){90}}
\put(60,-30){\line(0,1){80}}
\put(-150,50){\line(3,-1){250}}

\put(0,0){\circle*{5}}\put(0,40.5){\circle*{5}}\put(0,-19){\circle*{5}}
\put(60,40){\circle*{5}}

\put(60,-20){\circle{5}}

\put(115,5){\mbox{$h^2$}}
\put(-10,-42){\mbox{$h^2=0$}} \put(-22,-55){\mbox{\rm Macdonalds}}
\put(50,-42){\mbox{$h^2=\frac{1}{2}$}} \put(47,-55){\mbox{\rm 3-Schurs}}

\put(5,55){\mbox{$\lambda$}}

\linethickness{0.5mm}
\qbezier(-150,-20)(0,-19)(40,-18)\qbezier(40,-18)(50,-18)(40,-15)\qbezier(0,0)(30,-12)(40,-15)
\qbezier(-150,40)(-110,37)(0,0)
\qbezier(-150,50)(-135,45)(-120,43)  \qbezier(-120,43)(-60,41)(100,40)
\qbezier(100,-33.3)(40,-18)(100,-20)

\end{picture}

\noindent
This last picture corresponds to "unphysical" situation when $|\sin(3\theta)|>1$.

Coming back to the switch between branches at the points $\sin(3\theta)=0$ and
$\sin(3\theta)=\pm 1$, it is useful to redraw the pictures in still another projection:

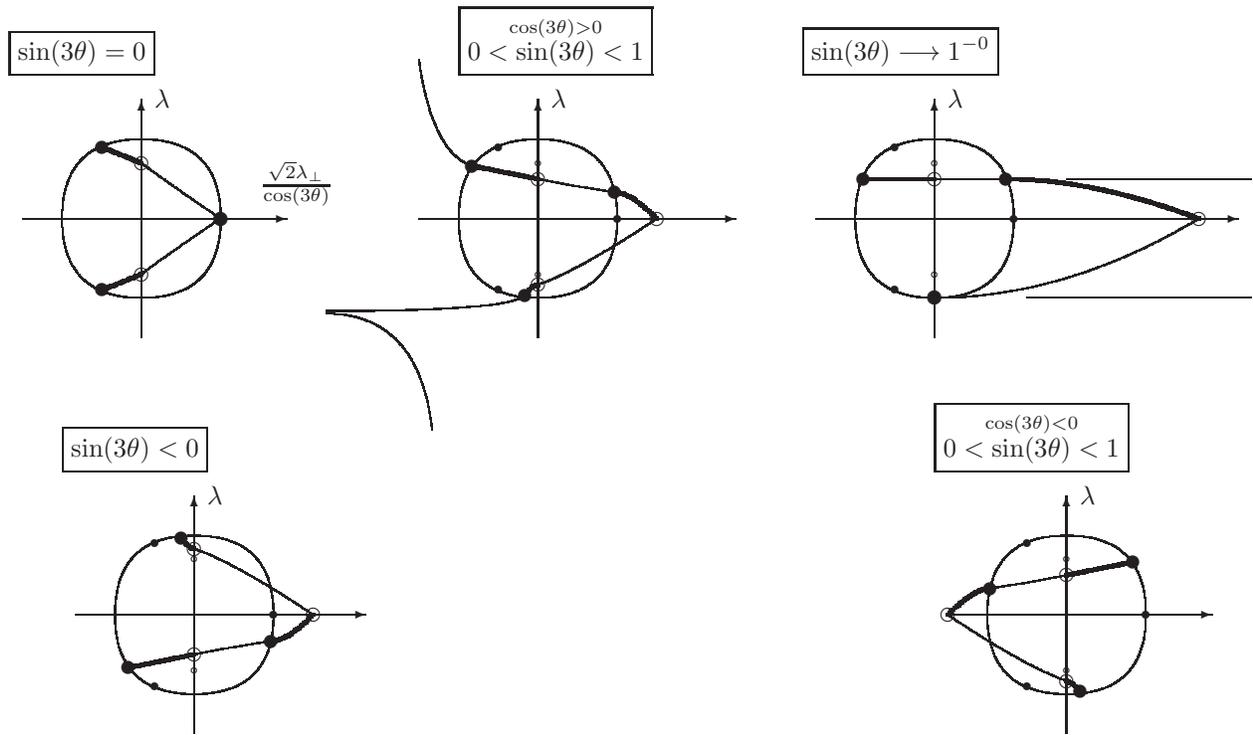
\begin{picture}(300,290)(-40,-200)

\put(-50,60){\mbox{$\boxed{\sin(3\theta)=0}$}}
\qbezier(-30,0)(-30,30)(0,30)\qbezier(0,30)(30,30)(30,0)
\qbezier(30,0)(30,-30)(0,-30)\qbezier(0,-30)(-30,-30)(-30,0)
\put(-45,0){\vector(1,0){100}}\put(45,10){\mbox{$\frac{\sqrt{2}\lambda_{\bot_{}}}{\cos(3\theta)}$}}
\put(0,-45){\vector(0,1){90}}\put(5,42){\mbox{$ \lambda $}}

\put(30,0){\circle*{5}}\put(-15,27){\circle*{5}}\put(-15,-27){\circle*{5}}
\put(30,0){\circle{5}}\put(0,21){\circle{5}}\put(0,-21){\circle{5}}
\qbezier(30,0)(15,10)(0,21)\qbezier(30,0)(15,-10)(0,-21)

 \put(150,0){
 \put(-30,60){\mbox{$\boxed{\stackrel{\cos(3\theta)>0}{0<\sin(3\theta)<1}}$}}
\qbezier(-30,0)(-30,30)(0,30)\qbezier(0,30)(30,30)(30,0)
\qbezier(30,0)(30,-30)(0,-30)\qbezier(0,-30)(-30,-30)(-30,0)
\put(-45,0){\vector(1,0){120}}
\put(0,-45){\vector(0,1){90}}\put(5,42){\mbox{$ \lambda $}}

\put(30,0){\circle*{3}}\put(-15,27){\circle*{3}}\put(-15,-27){\circle*{3}}
\put(30,0){\circle{2}}\put(0,21){\circle{2}}\put(0,-21){\circle{2}}
\put(29,10){\circle*{5}}\put(-25,20){\circle*{5}}\put(-5,-29){\circle*{5}}
\put(45,0){\circle{5}}\put(0,15){\circle{5}}\put(0,-25){\circle{5}}
\qbezier(45,0)(15,-20)(0,-25)\qbezier(29,10)(15,12)(0,15)

\qbezier(-25,20)(-40,25)(-45,60)
\qbezier(-5,-29)(-10,-35)(-80,-35)
\qbezier(-80,-36)(-45,-36)(-40,-80)

  \put(150,0){
  \put(-50,60){\mbox{$\boxed{\sin(3\theta)\longrightarrow 1^{-0}}$}}
  \qbezier(-30,0)(-30,30)(0,30)\qbezier(0,30)(30,30)(30,0)
\qbezier(30,0)(30,-30)(0,-30)\qbezier(0,-30)(-30,-30)(-30,0)
\put(-45,0){\vector(1,0){160}}
\put(0,-45){\vector(0,1){90}}\put(5,42){\mbox{$ \lambda $}}

\put(30,0){\circle*{3}}\put(-15,27){\circle*{3}}\put(-15,-27){\circle*{3}}
\put(30,0){\circle{2}}\put(0,21){\circle{2}}\put(0,-21){\circle{2}}
\put(27,15){\circle*{5}}\put(-27,15){\circle*{5}}\put(0,-30){\circle*{5}}
\put(100,0){\circle{5}}\put(0,15){\circle{5}}\put(0,-30){\circle{5}}
\qbezier(27,15)(10,15)(0,15)\qbezier(100,0)(50,-30)(0,-30)
\put(50,15){\line(1,0){70}}
\put(35,-30){\line(1,0){85}}

    \put(50,-150){
    \put(-50,60){\mbox{$\boxed{\stackrel{\cos(3\theta)<0}{0<\sin(3\theta)<1}}$}}
    \qbezier(-30,0)(-30,30)(0,30)\qbezier(0,30)(30,30)(30,0)
\qbezier(30,0)(30,-30)(0,-30)\qbezier(0,-30)(-30,-30)(-30,0)
\put(-45,0){\vector(1,0){100}}
\put(0,-45){\vector(0,1){90}}\put(5,42){\mbox{$ \lambda $}}

\put(30,0){\circle*{3}}\put(-15,27){\circle*{3}}\put(-15,-27){\circle*{3}}
\put(30,0){\circle{2}}\put(0,21){\circle{2}}\put(0,-21){\circle{2}}
\put(-29,10){\circle*{5}}\put(25,20){\circle*{5}}\put(5,-29){\circle*{5}}
\put(-45,0){\circle{5}}\put(0,15){\circle{5}}\put(0,-25){\circle{5}}
\qbezier(-45,0)(-15,-20)(0,-25)\qbezier(-29,10)(-15,12)(0,15)

\put(-330, 0){
    \put(-50,60){\mbox{$\boxed{ \sin(3\theta)<0 }$}}
\qbezier(-30,0)(-30,30)(0,30)\qbezier(0,30)(30,30)(30,0)
\qbezier(30,0)(30,-30)(0,-30)\qbezier(0,-30)(-30,-30)(-30,0)
\put(-45,0){\vector(1,0){110}}
\put(0,-45){\vector(0,1){90}}\put(5,42){\mbox{$ \lambda $}}

\put(30,0){\circle*{3}}\put(-15,-27){\circle*{3}}\put(-15,27){\circle*{3}}
\put(30,0){\circle{2}}\put(0,-21){\circle{2}}\put(0,21){\circle{2}}
\put(29,-10){\circle*{5}}\put(-25,-20){\circle*{5}}\put(-5,29){\circle*{5}}
\put(45,0){\circle{5}}\put(0,-15){\circle{5}}\put(0,25){\circle{5}}
\qbezier(45,0)(15,20)(0,25)\qbezier(29,-10)(15,-12)(0,-15)

\linethickness{0.5mm}

\qbezier(29,-10)(35,-10)(45,0)\qbezier(0,25)(-3,26)(-5,29)\qbezier(0,-15)(-15,-18)(-25,-20)
}

\linethickness{0.5mm}
\qbezier(-29,10)(-35,10)(-45,0)\qbezier(0,-25)(3,-26)(5,-29)\qbezier(0,15)(15,18)(25,20)
    }

\linethickness{0.5mm}
\qbezier(-27,15)(-10,15)(0,15)\qbezier(27,15)(60,15)(100,0)
  }

\linethickness{0.5mm}

\qbezier(29,10)(35,10)(45,0)\qbezier(0,-25)(-3,-26)(-5,-29)\qbezier(0,15)(-15,18)(-25,20)
}

\linethickness{0.5mm}
\qbezier(0,21)(-7,24)(-15,27)\qbezier(0,-21)(-7,-24)(-15,-27)

\end{picture}

\noindent
Coordinates on the plane are $\lambda$ and $\lambda_\bot$.
Thick lines show the evolution from $h^2=\frac{1}{2}$ (black dots) to $h^2=0$ (white dots).
Thin lines describe evolution in the region $h^2>\frac{1}{2}$, where some of the
lines/branches
actually merge.
Evolution in "unphysical" region $h^2<0$ is more involved and is schematically shown
in just one of the pictures.
The series of pictures clarifies how reshufflings at $\sin(3\theta)=0$
and $|\sin(3\theta)|=1$ take place.
In the former case the rightmost white and black dots merge with no essential reshuffling
of evolution lines.
In the latter case this white dot runs to $+\infty$ and re-appears at $-\infty$,
with an abrupt switch of roles between thick and thin segments of evolution lines.
Small circles show the positions of eigenvalues at $\theta=0$.
Those on the vertical line are $\lambda_\bot^\pm = \pm 1$ and maximal deviation from them
are to $|\lambda_\bot| = 2^{\pm 2}$, what corresponds to
$t^{\pm 1}=\frac{1+3q}{3+q}$ -- beyond this region one should consider complex angles
$\theta$.
This is also needed is $t$ and/or $q$ are complex.

\subsection{In the search for triality
}

In the limit $h\longrightarrow 0$ the three eigenvalues $\lambda$ tend
to $0$ and to the two values
$\sqrt{\frac{(1-q)(1+t)}{(1+q)(1-t)}}=i\sqrt{\frac{(1-q)(1+t^{-1})}{(1+q)(1-t^{-1})}}$
and
$-\sqrt{\frac{(1+q)(1-t)}{(1-q)(1+t)}}=i\sqrt{\frac{(1+q)(1-t^{-1})}{(1-q)(1+t^{-1})}}$
which are obviously related by the symmetry $q\longleftrightarrow t^{-1}$.
The same symmetry relates the two Macdonald polynomials if we write them as
\be
 M_2^+=
\frac{1}{2}\left(i\sqrt{\frac{(1-q)(1+t^{-1})}{(1+q)(1-t^{-1})}}\cdot P_2+p_1^2\right),
\ \ \ \ \
 M_2^- =
\frac{1}{2}\left(i\sqrt{\frac{(1+q)(1-t^{-1})}{(1-q)(1+t^{-1})}}\cdot P_2 + p_1^2\right)
\label{Maclevel2}
\ee
Note that the sign difference in the coefficients of $p_2$ is an automatic consequence
of this symmetry -- results from a transition from symmetric formulation with $t^{-1}$
to asymmetric, but more practical formulation with $t$.
The third Schur tends to a function of the "hidden" variable $\tilde P_2$
and decouples from the world of 2-Schurs, which depend on $P_2$ and $p_1$
(remaining orthogonal to them).

A natural question is what happens to the third 3-Schur functions in this limit
and if there is any sign of triality --
an expected symmetry between $q_1=q,\ q_2=t^{-1}$ and $q_3=tq^{-1}$.
Unfortunately, direct association with three "original" values of $\theta$,
is not simple: characteristic equations involve
\be
\sin(3\theta) =
-4\sin\theta\cdot\sin\left(\theta+\frac{2\pi}{3}\right)\cdot\sin\left(\theta-\frac{2\pi}{3}\right)
= 2\sqrt{2}\frac{q-t}{\sqrt{(1-q^2)(1-t^2)}}
\ee
and are fully $Z_3$-invariant.
Resolution of triality mystery remains among the challenges for the future work.
One of the inspirations here can be that the third parameter $tq^{-1}$ naturally appears
in the cut-and-join operator for {\it generalized} Macdonald functions of \cite{genpols} --
and this can be right direction to look for triality. 

\subsection{What happened: a summary}

We  now summarize what actually happened in this simple level-two example.
We started from a $Z_3$-symmetric \underline{triple} of 3-Schur functions,
and rescaled with the help of $h$-deformation one of the \underline{two} dimensions
in the  $(p_2,\tilde p_2)$ \underline{plane} --
this led to the \underline{pair} of ordinary 2-Schur functions,
while the \underline{third} one "decoupled".
If we do the same,
but first $\underline{\theta}$-rotated the $(p_2,\tilde p_2)$ \underline{plane},
then the same procedure leads to a \underline{pair} of Macdonald functions.
The reason why deviation of $\theta$ from zero affects the answer is that
the symmetry in the $(p_2,\tilde p_2)$ \underline{plane} was just a $Z_3\subset SO(2)$,
not the full rotation group: thus rescaling "along" one of the preferred directions
(associated with one of the special angles $\theta = 0, \pm\frac{2\pi}{3}$
and with particular 3-Schur function) is different from that in arbitrary direction.
As a remnant of original $Z_3$-symmetry we get triality --
when parameters $q_{1,2,3}$ are changed under of a simple rotation.

Underlined in the above paragraph are  numbers, peculiar to the level \underline{two},
at other levels they change -- but the whole scheme remains the same.

\section{Level 3}

\subsection{Macdonald polynomials}

In coordinates, which respect (\ref{metricforMac}), the three
Macdonald polynomials at level 3, complementing (\ref{Maclevel2}),
\be
M_{[1]}=p_1 \nn \\
M_{[2]}^+= M_{[2]} =
\frac{1}{2}\left(\sqrt{\frac{(1-q)(1+t )}{(1+q)(1-t )}}\cdot p_2+p_1^2\right),
\ \ \ \ \
M_{[2]}^- = M_{[11]}=
\frac{1}{2}\left(-\sqrt{\frac{(1+q)(1-t )}{(1-q)(1+t )}}\cdot p_2 + p_1^2\right)
\label{Maclevel2a}
\ee
are:
\be
M_{[3]}^+=M_{[3]}=\frac{(1-q)\sqrt{1+t+t^2}}{(1-t)\sqrt{1+q+q^2}}\cdot\frac{p_3}{3}
+ \sqrt{\frac{(1-q)(1+t)}{(1+q)(1-t)}}\cdot\frac{p_2p_1}{2} + \frac{p_1^3}{6}
\nn \\
M_{[3]}^-=M_{[111]}=
\frac{(1-t)\sqrt{1+q+q^2}}{(1-q)\sqrt{1+t+t^2}}\cdot\frac{p_3}{3}
- \sqrt{\frac{(1+q)(1-t)}{(1-q)(1+t)}}\cdot\frac{p_2p_1}{2} + \frac{p_1^3}{6} \nn \\
M_{[21]}^0 =M_{[21]}= -\sqrt{(1+q+q^2)(1+t+t^2)}\cdot\frac{p_3}{3}
+ (t-q)\sqrt{\frac{(1+q)(1+t)}{(1-q)(1-t)}}\cdot\frac{p_2p_1}{2}
+ (2+q+t+2qt)\cdot\frac{p_1^3}{6}
\label{Maclevel3}
\ee
The squared norms of these polynomials in metric (\ref{metricforMac}) are:
\be
||M_{[1]}||^2=1, \ \ \ \ \
||M_{[2]}^+||^2 = \frac{1-qt}{(1+q)(1-t)}, \ \ \ \
||M_{[2]}^-||^2 = \frac{1-qt}{(1-q)(1+t)}, \nn \\
||M_{[3]}^+||^2 = \frac{(1-qt)(1-q^2t)}{(1+q)(1+q+q^2)(1-t)^2}, \ \ \ \
||M_{[3]}^-||^2 = \frac{(1-qt)(1-q^2t)}{(1-q)^2(1+t)(1+t+t^2)}, \nn\\
||M_{[21]}||^2 = \frac{(1-qt^2)(1-q^2t)}{(1-q)(1-t)}
\ee

Extension of operator (\ref{WM2level2}) is
\be
\widehat{WM}_{[2]}=\frac{p_2}{2}\p_1^2 + p_1^2\p_2 -  {2\sigma p_2\p_2}
+ u\cdot (2p_3\p_2\p_1+3p_2p_1\p_3) + v\cdot p_3\p_3+w\cdot p_2p_1\p_2\p_1
+ \ldots,
\label{WM2level3}
\ee
where
\be
u=\frac{2\sqrt{(1+q+q^2)(1+t+t^2)}}{2+q+t+2qt} \ \stackrel{t=q}{\longrightarrow} \ 1
\ee
describes the $q,t$-deformation of a term, present in the ordinary (2-Schur)
cut-and-join operator, while the three other parameters
are peculiar for Macdonald polynomials (and vanish for 2-Schurs):
\be
\sigma =  \frac{q-t }{\sqrt{(1-q^2)(1-t^2)}}, \ \ \
v=-\frac{9(q-t)}{2+q+t+2qt}\,\sqrt{\frac{(1+q)(1+t)}{(1-q)(1-t)}}, \ \ \
w = - \frac{2(q-t)}{2+q+t+2qt}\,\sqrt{\frac{(1-q)(1-t)}{(1+q)(1+t}}
\ee
In addition to this $\widehat{WM}_{[2]}$ at level 3 we can observe two higher $W$-operators
$\widehat{WM}_{[3]}$ and $\widehat{WM}_{[21]}$, which were vanishing at level two.

\subsection{3-Schur functions}

According to \cite{3Schur} the six Schur functions associated with six plane partitions
 at level three (i.e. made from three cubes)
\be
{\cal S}_{[3]}^0 = \frac{2\tilde{\tilde p}_3
+ \sqrt{6}\tilde p_3}{3}+\frac{\tilde p_2p_1}{\sqrt{2}}
+\frac{p_1^3}{6}, \ \ \ \
{\cal S}_{[3]}^\pm = \frac{4\tilde{\tilde p}_3-\sqrt{6}\tilde p_3\pm 3\sqrt{2}  p_3}{6}
+\frac{(- \tilde p_2\pm \sqrt{3 }  p_2)p_1}{2\sqrt{2}}
+\frac{p_1^3}{6},
\nn \\
{\cal S}_{[21]}^0=  \frac{-\tilde{\tilde p}_3+ \sqrt{6}\tilde p_3}{6}
-\frac{\tilde p_2p_1}{\sqrt{2}}  +\frac{p_1^3}{3}, \ \ \ \
{\cal S}_{[21]}^\pm= \frac{-2\tilde{\tilde p}_3-\sqrt{6}\tilde p_3\pm 3\sqrt{2} p_3}{12}
-\frac{(- \tilde p_2\pm \sqrt{3 }p_2 )p_1}{2\sqrt{2}}    +\frac{p_1^3}{3}
\label{3Schurlevel3}
\ee
Time variables $\vec p_2 = (p_2,\tilde p_2)$ are the same as at level two.
Now they are complemented by a three-dimensional vector
$\vec p_3 = (p_3,\tilde p_3, \tilde{\tilde p}_3)$.
The Cauchy formula persists:
\be
\sum_\pi \frac{{\cal S}_\pi\{p\}{\cal S}_\pi\{p'\}}{<{\cal S}_\pi|{\cal S}_\pi>} =
\exp\left(\sum_n \frac{\vec p_n\vec p_n\!'}{n}\right) =
\exp\left(p_1p_1'+\frac{p_2p_2'+\tilde p_2\tilde p_2'}{2}
+\frac{ {p_3p_3'} +\tilde p_3\tilde p_3'+\tilde{\tilde p}_3\tilde{\tilde p}_3'}{3}
+ \ldots\right)
\label{Cauchy3}
\ee
The simplest cut-and-join operator is extended to
\be
\hat{\cal W}_{[2]}^0 = \frac{  p_2}{2}\frac{\p^2}{\p p_1^2}+p_1^2\frac{\p}{\p   p_2}
 -\frac{1}{\sqrt{2}}\Big(\tilde p_2\frac{\p}{\p p_2} + p_2\frac{\p}{\p \tilde p_2}\Big)
+\frac{3(\sqrt{3}   p_3+   p_2 p_1)}{2}\frac{\p}{\p \tilde{\tilde p}_3}
-\frac{3(   p_3+\sqrt{3}   p_2 p_1)}{2\sqrt{2}}\frac{\p}{\p \tilde p_3} +
\nn \\
+\frac{3}{2}\left(\sqrt{3} \tilde{\tilde p}_3
-\frac{\tilde p_3 + \sqrt{3} \tilde p_2 p_1)}{\sqrt{2}}\right)
\frac{\p}{\p p_3}
+\frac{ -\sqrt{3} p_3+ p_2 p_1}{\sqrt{2}}\frac{\p^2}{\p \tilde p_2 \p p_1}
+\left(\tilde{\tilde p}_3+\frac{-\sqrt{3}\tilde p_3+ \tilde p_2 p_1}{\sqrt{2}}\right)
\frac{\p^2}{\p  p_2\p p_1} + \ldots
\label{W203}
\ee
(as usual, dots stand for terms which do not act at level three and will be omitted)
and  it now acquires six new eigenfunctions (\ref{3Schurlevel3}):
\be
\hat{\cal W}_{[2]}^0 {\cal S}_{[3]}^0 = 0, \ \ \ \
\hat{\cal W}_{[2]}^0 {\cal S}_{[3]}^\pm = \pm 3\sqrt{\frac{3}{2}}{\cal S}_{[3]}^\pm,
\ \ \ \ \ \ \ \ \ \
\hat{\cal W}_{[2]}^0 {\cal S}_{[21]}^0 = 0, \ \ \ \
\hat{\cal W}_{[2]}^0 {\cal S}_{[21]}^\pm = \mp \frac{3}{2}\sqrt{\frac{3}{2}}{\cal
S}_{[21]}^\pm
\ee
Also extended is (\ref{W20bot}):
$$
\hat{\cal W}_{[2]}^{0,\bot} =
\frac{\tilde  p_2}{2}\frac{\p^2}{\p p_1^2}+p_1^2\frac{\p}{\p \tilde  p_2}
+ \frac{1}{\sqrt{2}}\Big( p_2\frac{\p}{\p p_2} - \tilde p_2\frac{\p}{\p \tilde p_2}\Big)
\Big(p_1\frac{\p}{\p p_1}-1\Big)
+ \frac{3}{2\sqrt{2}}\Big( \tilde p_3\frac{\p}{\p \tilde p_3} - p_3\frac{\p}{\p p_3}\Big)
+ \frac{3\sqrt{3}}{2}
\Big(\tilde{\tilde p}_3\frac{\p}{\p \tilde p_3}+\tilde p_3\frac{\p}{\p\tilde{\tilde
p}_3}\Big)
+
$$
\vspace{-0.4cm}
\be
+ \frac{1}{2}\Big(3\tilde p_2p_1\frac{\p}{\p \tilde{\tilde p}_3}+
2\tilde{\tilde p}_3\frac{\p^2}{\p \tilde p_2\p p_1}\Big)
+  \frac{\sqrt{3}}{2\sqrt{2}} \left(\Big(3\tilde p_2p_1\frac{\p}{\p \tilde p_3}+
2\tilde p_3\frac{\p^2}{\p \tilde p_2\p p_1}\Big)   -
\Big(3p_2p_1\frac{\p}{\p p_3}+
2 p_3\frac{\p^2}{\p  p_2\p p_1}\Big)\right) + \ldots
\ee
with
\be
\hat{\cal W}_{[2]}^{0,\bot} {\cal S}_{[3]}^0 = 3\sqrt{2}\,{\cal S}_{[3]}^0, \ \ \ \
\hat{\cal W}_{[2]}^{0,\bot}{\cal S}_{[3]}^\pm =  -\frac{3}{\sqrt{2}}{\cal S}_{[3]}^\pm,
\ \ \ \ \ \ \ \ \ \
\hat{\cal W}_{[2]}^{0,\bot} {\cal S}_{[21]}^0 = -\frac{3}{\sqrt{2}}{\cal S}_{[21]}^0, \ \ \
\
\hat{\cal W}_{[2]}^{0,\bot} {\cal S}_{[21]}^\pm = \frac{3}{2\sqrt{2}}\,{\cal S}_{[21]}^\pm
\ee
as well as the raising and lowering operators $\hat{\cal W}_{[2]}^{\pm}$ and
$\hat{\cal W}_{[2]}^{\pm,\bot}$.

\subsection{Level-three cut-and-join operators}

Also emerging at level three is the whole new family $\hat{\cal W}_{[3]}$,
which were not seen (acted by zero) at lower levels.
In particular there is a triple (since the $\vec p_3$ is now three-dimensional)
of linear independent "Cartanian" operators:
$$
\hat{\cal W}_{[3]}^0 = \Big(p_3\frac{\p^3}{\p p_1^3} + 3p_1^3\frac{\p}{\p p_3}\Big)
- 3\sqrt{\frac{3}{2}}\Big(p_2p_1\frac{\p^2}{\p \tilde p_2\p p_1}
+ \tilde p_2p_1\frac{\p^2}{\p p_2\p p_1}\Big)
+ \frac{15}{2}\Big(\tilde{\tilde p}_3\frac{\p}{\p p_3} + p_3\frac{\p}{\p\tilde{\tilde
p}_3}\Big)
-\frac{9}{2}\sqrt{\frac{3}{2}}\Big(p_3\frac{\p}{\p \tilde p_3} + \tilde p_3\frac{\p}{\p
p_3}\Big)
+
$$
\vspace{-0.4cm}
\be
+ \frac{3\sqrt{3}}{2}\Big(3p_2p_1\frac{\p}{\p \tilde{\tilde p}_3}+
2\tilde{\tilde p}_3\frac{\p^2}{\p p_2\p p_1} \Big)
- \frac{3}{2\sqrt{2}}\Big(3p_2p_1\frac{\p}{\p {\tilde p}_3}+
2{\tilde p}_3\frac{\p^2}{\p p_2\p p_1} \Big)
- \frac{3}{2\sqrt{2}}\Big(3\tilde p_2p_1\frac{\p}{\p { p}_3}+
2 p_3\frac{\p^2}{\p \tilde p_2\p p_1} \Big)
\ee
with
\be
\hat{\cal W}_{[3]}^{0 } {\cal S}_{[3]}^0 = 0, \ \ \ \
\hat{\cal W}_{[3]}^{0 }{\cal S}_{[3]}^\pm =  \pm 9\sqrt{2}\cdot {\cal S}_{[3]}^\pm,
\ \ \ \ \ \ \ \ \ \
\hat{\cal W}_{[3]}^{0 } {\cal S}_{[21]}^0 = 0, \ \ \ \
\hat{\cal W}_{[3]}^{0 } {\cal S}_{[21]}^\pm = \pm\frac{9}{2\sqrt{2}}\,{\cal S}_{[21]}^\pm
\ee
then
$$
\hat{\cal W}_{[3]}^{0,\bot} =
\Big(\tilde p_3\frac{\p^3}{\p p_1^3} + 3p_1^3\frac{\p}{\p \tilde p_3}\Big)
+ 3\sqrt{\frac{3}{2}}\Big(\tilde p_2p_1\frac{\p^2}{\p \tilde p_2\p p_1}
- p_2p_1\frac{\p^2}{\p p_2\p p_1}\Big)
+ \frac{15}{2}\Big(\tilde{\tilde p}_3\frac{\p}{\p \tilde p_3}
+ \tilde p_3\frac{\p}{\p\tilde{\tilde p}_3}\Big)
+\frac{9}{2}\sqrt{\frac{3}{2}}\Big(\tilde p_3\frac{\p}{\p \tilde p_3}
-  p_3\frac{\p}{\p p_3}\Big)
+
$$
\vspace{-0.4cm}
\be
+ \frac{3\sqrt{3}}{2}\Big(3\tilde p_2p_1\frac{\p}{\p \tilde{\tilde p}_3}+
2\tilde{\tilde p}_3\frac{\p^2}{\p \tilde p_2\p p_1} \Big)
+\frac{3}{2\sqrt{2}}\Big(3\tilde p_2p_1\frac{\p}{\p {\tilde p}_3}+
2{\tilde p}_3\frac{\p^2}{\p \tilde p_2\p p_1} \Big)
- \frac{3}{2\sqrt{2}}\Big(3 p_2p_1\frac{\p}{\p { p}_3}+
2 p_3\frac{\p^2}{\p  p_2\p p_1} \Big)
\ee
with
\be
\hat{\cal W}_{[3]}^{0,\bot } {\cal S}_{[3]}^0 = 6\sqrt{6}\cdot {\cal S}_{[3]}^0 , \ \
\hat{\cal W}_{[3]}^{0,\bot }{\cal S}_{[3]}^\pm = -3\sqrt{6}\cdot {\cal S}_{[3]}^\pm,
\ \  \ \
\hat{\cal W}_{[3]}^{0,\bot } {\cal S}_{[21]}^0 = 3\sqrt{\frac{3}{2}}\cdot {\cal S}_{[21]}^0,
\ \
\hat{\cal W}_{[3]}^{0,\bot } {\cal S}_{[21]}^\pm
= -\frac{3}{2}\sqrt{\frac{3}{2}}\cdot{\cal S}_{[21]}^\pm
\ee
and
$$
\hat{\cal W}_{[3]}^{0,\bot\bot} =
\Big(\tilde{\tilde p}_3\frac{\p^3}{\p p_1^3} + 3p_1^3\frac{\p}{\p \tilde{\tilde p}_3}\Big)
+3\Big(\tilde p_2p_1\frac{\p^2}{\p \tilde p_2\p p_1}
+  p_2p_1\frac{\p^2}{\p p_2\p p_1}\Big)
+ \frac{15}{2}\Big({\tilde p}_3\frac{\p}{\p \tilde p_3} + p_3\frac{\p}{\p  p_3}\Big)
+ \frac{21}{2}\,\tilde{\tilde p}_3\frac{\p}{\p \tilde{\tilde p}_3}
+
$$
\vspace{-0.4cm}
\be
+ \frac{3\sqrt{3}}{2} \Big(3\tilde p_2p_1\frac{\p}{\p {\tilde p}_3}+
2{\tilde p}_3\frac{\p^2}{\p \tilde p_2\p p_1} \Big)
+ \frac{3\sqrt{3}}{2}\Big(3 p_2p_1\frac{\p}{\p { p}_3}+
2 p_3\frac{\p^2}{\p  p_2\p p_1} \Big)
\ee
with
\be
\hat{\cal W}_{[3]}^{0,\bot\bot } {\cal S}_{[3]}^0 = 12{\cal S}_{[3]}^0, \ \ \ \
\hat{\cal W}_{[3]}^{0,\bot\bot }{\cal S}_{[3]}^\pm =  12 {\cal S}_{[3]}^\pm,
\ \ \ \ \ \ \ \ \ \
\hat{\cal W}_{[3]}^{0,\bot\bot} {\cal S}_{[21]}^0 = -\frac{3}{2}\,{\cal S}_{[21]}^0, \ \ \ \
\hat{\cal W}_{[3]}^{0,\bot\bot } {\cal S}_{[21]}^\pm = -\frac{3}{2}\,{\cal S}_{[21]}^\pm
\ee
Note that there is no full symmetry between the three times $p_3$, $\tilde p_3$
and $\tilde{\tilde p}_3$
-- like there was no between $p_2$ and $\tilde p_2$ already at level 2.
Still in the latter case the symmetry was just $Z_3$, while for $\vec p_3$ it is a little
more involved.

Like the linear independent pair ($\hat{\cal W}_{[2]}^0$, $\hat{\cal W}_{[2]}^{0,bot}$)
was treated in \cite{3Schur} as an explicitly symmetric (but not independent) triple
$\hat{\cal W}_{[2]}^0$, $\hat{\cal W}_{[2]}^{0'}$
$\hat{\cal W}_{[2]}^{0''}$ one can do something similar with $W_{[3]}$ family.
An option is to label such {\it overfull} basis of Cartanian operators by plane partitions.
The study of these structures, as well as of additional "raising" and "lowering"
${\cal W}$-operators, which can make it truly "rigid", is beyond the scope of the present
paper.

\subsection{Interpolation from 3-Schurs to 2-Schurs}

Interpolation to the ordinary 2-Schurs at level three is rather straightforward.
First of all, we assume that evolution affects not the scalar product
(\ref{metricfor3Schur}), but the shape of functions and operators of the
given time-variables -- perhaps, by making $h$-dependent rotations and rescalings.
This means that evolution preserves hermiticity: self-conjugate combinations
remain self-conjugate.
Thus it is convenient to rewrite (\ref{W203}),
putting together the self-conjugate combinations
(as we already did for more complicated ${\cal W}$-operators in the previous subsection):
\be
\hat{\cal W}_{[2]}^0  =\Big( \frac{  p_2}{2}\frac{\p^2}{\p p_1^2}+p_1^2\frac{\p}{\p
p_2}\Big)
+\frac{1}{2}\Big(2\tilde{\tilde p}_3 \frac{\p^2}{\p  p_2\p p_1}
+ 3p_2p_1\frac{\p}{\p \tilde{\tilde p}_3}\Big)
- \frac{1}{2}\sqrt{\frac{3}{2}}
\Big(2\tilde p_3\frac{\p^2}{\p  p_2\p p_1}+ 3p_2p_1\frac{\p}{\p \tilde p_3}\Big)+
\label{W203b}
\ee
{\footnotesize
$$
+\frac{1}{\sqrt{2}}
\Big(\tilde p_2\frac{\p}{\p p_2} + p_2\frac{\p}{\p \tilde p_2}\Big)\,
\Big(p_1\frac{\p}{\p p_1}-1\Big)
+\frac{3\sqrt{3}}{2}\Big(p_3\frac{\p}{\p \tilde{\tilde p}_3}
+ \tilde{\tilde p}_3\frac{\p}{\p p_3}\Big)
- \frac{3}{2\sqrt{2}}\Big(p_3\frac{\p}{\p \tilde p_3} + \tilde p_3\frac{\p}{\p p_3}\Big)
-\frac{\sqrt{3}}{2\sqrt{2}}
\Big(2p_3\frac{\p^2}{\p \tilde p_2\p p_1}   + 3 \tilde p_2 p_1\frac{\p}{\p p_3}\Big)
$$
}
and compare this with the specialization of  (\ref{WM2level3}) to $t=q$,
\be
\hat{W}_{[2]}=\frac{p_2}{2}\p_1^2 + p_1^2\p_2
+  \Big(2\P_3 \frac{\p^2}{\p p_2\p p_1}+3p_2p_1\frac{\p}{\p\P_3}\Big)
+ \ldots
\label{WS2level3}
\ee
Clearly, the $h$-evolution to $h=0$ should eliminate all the terms
in the second line of (\ref{W203b}).
We introduced here a new notation $\P_3$, because the time variable,
surviving after  reduction, is not going to be just our $p_3$.
As to the first line in (\ref{W203b}), it looks appealing to just say that this
{\it surviving} variable is
$\P_3 \ \stackrel{?}{\sim}\  \tilde{\tilde p}_3-\sqrt{\frac{3}{2}}\,\tilde p_3$.
However, in (\ref{3Schurlevel3}), that ${\cal S}^0_{[3]}$ and ${\cal S}_{[21]}^\pm$
which should decouple  at the 2-Schur point $h=0$, rather depend on
$\P_3^\bot \ \stackrel{?}{\sim}\  \tilde{\tilde p}_3+\sqrt{\frac{3}{2}}\,\tilde p_3$,
which is {\it not} orthogonal to {\it such} $\P_3$.
Still, as emphasized in \cite{3Schur},
the fact that all the three depend on just one such combination,
(as well as on a straightforwardly orthogonal $p_3$),
is a {\it miracle}, which makes a straightforward reduction possible --
but now we need more technical details on how it works.
Fortunately, our would-be $\P_3$ and $\P_3^\bot$ get orthogonal at $h=0$,
provided we substitute  $\sqrt{\frac{3}{2}}\longrightarrow \sqrt{1+h^2}$ in above
expressions
--
as we once did in the $\vec p_2$ sector at level two.
Still, we can not keep them orthogonal for all $h$ -- and at level three there is no
longer {\it ideal} choice of coordinates, suitable for the $h$-evolution.
The choice in \cite{3Schur} was $h$-independent, orthogonal and properly normalized
\be
\P_3 :=  \sqrt{\frac{3}{5}}\tilde{\tilde p}_3 - \sqrt{\frac{2}{5}}\tilde p_3,
\ \ \ \ \ \ \ \ \
\P_3^\bot :=  \sqrt{\frac{2}{5}} \tilde{\tilde p}_3 + \sqrt{\frac{3}{5}}\tilde p_3
\ee
The price to pay is that in these variables
the operators $\hat {\cal W}(h)$ at $h\neq 0$  will have an admixture of $\P_3^\bot$.
Instead separation of variables is nice in Schur functions: at $h^2=\frac{1}{2}$
\be
{\cal S}_{[3]}^0 = \frac{2\tilde{\tilde p}_3
+ \sqrt{6}\tilde p_3}{3}+\frac{\tilde p_2p_1}{\sqrt{2}} +\frac{p_1^3}{6}
&= \frac{2}{3}\sqrt{\frac{5}{2}}\P_3^\bot +\frac{\tilde p_2p_1}{\sqrt{2}}+\frac{p_1^3}{6}
\nn \\
{\cal S}_{[3]}^\pm = \frac{4\tilde{\tilde p}_3-\sqrt{6}\tilde p_3\pm 3\sqrt{2}  p_3}{6}
+\frac{(- \tilde p_2\pm \sqrt{3 }  p_2)p_1}{2\sqrt{2}}+\frac{p_1^3}{6}
&= \frac{1}{6}\sqrt{\frac{2}{5}}\P_3^\bot + \sqrt{\frac{3}{5}}\P_3 \pm \frac{1}{\sqrt{2}}p_3
-\frac{\tilde p_2 p_1}{2\sqrt{2}}\pm \frac{1}{2}\sqrt{\frac{3}{2}}p_2p_1+\frac{p_1^3}{6}
\nn \\
{\cal S}_{[21]}^0=  \frac{-\tilde{\tilde p}_3+ \sqrt{6}\tilde p_3}{6}
-\frac{\tilde p_2p_1}{\sqrt{2}}  +\frac{p_1^3}{3}
&= \frac{1}{3}\sqrt{\frac{2}{5}}\P_3^\bot-\frac{1}{2}\sqrt{\frac{3}{5}}\P_3
-\frac{\tilde p_2p_1}{\sqrt{2}}  +\frac{p_1^3}{3}
\nn \\
{\cal S}_{[21]}^\pm= \frac{-2\tilde{\tilde p}_3-\sqrt{6}\tilde p_3\pm 3\sqrt{2} p_3}{12}
-\frac{(- \tilde p_2\pm \sqrt{3 }p_2 )p_1}{2\sqrt{2}}    +\frac{p_1^3}{3}
&= -\frac{1}{6}\sqrt{\frac{5}{2}}\P_3^\bot \pm \frac{p_3}{2\sqrt{2}}
+\frac{\tilde p_2 p_1}{2\sqrt{2}}\mp \frac{1}{2}\sqrt{\frac{3}{2}}p_2p_1 +\frac{p_1^3}{3}
\nn
\label{3Schurlevel3'}
\ee
and deformation to generic $h$ is:
\be
\begin{array}{rcl}
\nn\\
{\rm 3-Schur\ point} \  h^2=\frac{1}{2} \ \ \ \ \ \ \
&&  \ \ \ \ \ {\rm generic} \ h  \nn \\   \nn \\ \hline \nn \\
{\cal S}_{[3]}^0  =
\frac{2}{3}\sqrt{\frac{5}{2}}\P_3^\bot +\frac{\tilde p_2p_1}{\sqrt{2}}+\frac{p_1^3}{6}
&\longrightarrow
& \boxed{
\frac{1}{h}\cdot\left(\sqrt{\frac{3-h^2}{3h^2-1}}\cdot \frac{\P_3^\bot}{3}
+\frac{\tilde p_2p_1}{2}\right)
}
+\frac{p_1^3}{6}
\nn \\ \nn \\
\!\!\!\!\!\!\!\!\!\!\!\!\!\!\!\!\!\!\!\!\!\!\!\!\!\!\!\!\!
\boxed{{\cal S}_{[3]}^\pm}  =
 \frac{1}{6}\sqrt{\frac{2}{5}}\P_3^\bot +  \sqrt{\frac{3}{5}}\P_3 \pm \frac{1}{\sqrt{2}}p_3
-\frac{\tilde p_2 p_1}{2\sqrt{2}}\pm \frac{1}{2}\sqrt{\frac{3}{2}}p_2p_1+\frac{p_1^3}{6}
&\longrightarrow
&
\boxed{
 \frac{ 1+h^2}{ \sqrt{3 (1-h^2) (3-h^2)}}\cdot  \P_3 \pm \frac{\sqrt{1+h^2}}{2}\,p_2p_1
 +\frac{p_1^3}{6}
} \ + \nn \\ \nn \\
&&\ \ \ \ \ \ \ \ \ \ \ \ \ \ \ \ \ \ \ \ + h\cdot\left(
\sqrt{\frac{3h^2-1}{3-h^2}} \cdot \frac{\P_3^\bot}{3}  -\frac{\tilde p_2 p_1}{2 }
\pm \sqrt{\frac{(1+h^2)}{3(1-h^2)}}\cdot  p_3 \right)
\nn \\ \nn \\
\boxed{{\cal S}_{[21]}^0}
= \frac{1}{3}\sqrt{\frac{2}{5}}\P_3^\bot- \sqrt{\frac{3}{5}}\frac{\P_3}{2}
-\frac{\tilde p_2p_1}{\sqrt{2}}  +\frac{p_1^3}{3}
&\longrightarrow
&
\boxed{
-\sqrt{\frac{3 (1-h^2)}{ (3-h^2)}}\cdot   \frac{(1+h^2)\P_3}{3} +\frac{p_1^3}{3}
}
+ 2h\cdot\left( \sqrt{\frac{3h^2-1}{3-h^2}}\cdot \frac{\P_3^\bot}{3}
-\frac{\tilde p_2p_1}{2} \right)
\nn \\ \nn \\
\!\!\!\!\!\!\!\!\!\!\!\!\!\!\!\!\!\!\!\!\!\!\!\!\!
{\cal S}_{[21]}^\pm(h) = -\frac{1}{6}\sqrt{\frac{5}{2}}\P_3^\bot \pm \frac{p_3}{2\sqrt{2}}
+\frac{\tilde p_2 p_1}{2\sqrt{2}}\mp \frac{1}{2}\sqrt{\frac{3}{2}}p_2p_1 +\frac{p_1^3}{3}
& \longrightarrow
& \boxed{
\frac{1}{2h}\cdot \left(- {\sqrt{(3h^2-1)(3-h^2)}}\cdot \frac{\P_3^\bot}{3}
+{(1-h^2)} \tilde p_2 p_1 \pm \sqrt{\frac{1-h^4}{3}} p_3\right)
}\  +
\nn \\ \nn \\
&& \ \ \ \ \ \ \ \ \ \ \ \ \ \ \ \ \ \ \ \ \ \ \ \ \ \ \ \ \ \ \ \ \ \ \ \
 +\left(\frac{p_1^3}{3} \mp  \frac{\sqrt{1+h^2}}{2}\,p_2p_1\right)
\end{array}
\label{3Schurlevel3''}
\ee
Surviving in the 2-Schur limit, i.e. when $h=0$ are
the expressions in boxes.
As necessary, $S_{[3]}^\pm$ and $S_{[21]}^0$ are finite and reproduce
the standard 2-Schur functions, while $S_{[3]}^0$ and $S_{[21]}^\pm$ get singular,
but the leading pieces are independent of $\P_3,p_2$ and $p_1$ --
only on the "decoupling" time-variables $\P_3^\bot, p_3$ and $\tilde p_2$.
Actually, in this limit the coefficient in front of $\P_3^\bot$ becomes pure imaginary,
but this does not seem to cause any problem.
The $h$-dependence in the $\vec p_3$ sector can look somewhat sophisticated,
but it is unambiguously defined (deduced)
from that in the $\vec p_2$ sector and orthogonality,
once one fixes recursion rules in a nearly obvious way, see \cite{3Schur}.

The $h$-deformation of operator (\ref{W203b}) is
\be
\hat{\cal W}_{[2]}^0(h)  =\Big( \frac{  p_2}{2}\frac{\p^2}{\p p_1^2}+p_1^2\frac{\p}{\p
p_2}\Big)
+\sqrt{\frac{3(1-h^2)}{3-h^2}}\Big(2\P_3 \frac{\p^2}{\p  p_2\p p_1}
+ 3p_2p_1\frac{\p}{\p \P_3}\Big) -
\nn \\
- \frac{h}{2}\sqrt{\frac{3h^2-1}{3-h^2}}
\Big(2\P_3^\bot\frac{\p^2}{\p  p_2\p p_1}+ 3p_2p_1\frac{\p}{\p \P_3^\bot}\Big)+
+\underline{h\Big(\tilde p_2\frac{\p}{\p p_2} + p_2\frac{\p}{\p \tilde p_2}\Big)\,
\Big(p_1\frac{\p}{\p p_1}-1\Big)}+
\frac{6h}{\sqrt{1+h^2}}\Big(p_3\frac{\p}{\p \P_3}+ \P_3\frac{\p}{\p p_3}\Big)+
\nn
\ee
\vspace{-0.3cm}
\be
+ \frac{3}{2}\sqrt{\frac{3(1-h^2)(3h^2-1)}{3-h^2}}
\Big(p_3\frac{\p}{\p \P_3^\bot}+ \P_3^\bot\frac{\p}{\p p_3}\Big)
-\frac{\sqrt{3(1-h^2)}}{2}
\Big(2p_3\frac{\p^2}{\p \tilde p_2\p p_1}   + 3 \tilde p_2 p_1\frac{\p}{\p p_3}\Big)
\label{W203bh}
\ee
At $h=0$ the first line turns into (\ref{WS2level3}), as expected and the second line
vanishes.
However, the third line remains finite -- but it vanishes on functions of the variables
$\P_3, p_2, p1$, and in this sense the operator is properly reduced to the ordinary
2-Schur one.
Note also, that underlined term is occasionally zero at level $3$ (where there are just
two monomial on which it couild act, $\tilde p_2p1$ and $p_2p1$, which are both
of degree one in $p_1$), but it acts non-trivially at level $2$.
The eigenvalues of (\ref{W203bh}) for the $h$-deformed 3-Schur functions are:
\be
\hat{\cal W}_{[2]}^0(h) {\cal S}_{[2]}^0(h) = 0, \ \ \ \
\hat{\cal W}_{[2]}^0(h){\cal S}_{[2]}^\pm(h) =  \pm\sqrt{1+h^2}\,{\cal S}_{[2]}^\pm(h),\nn
\\
\hat{\cal W}_{[2]}^0(h) {\cal S}_{[3]}^0(h) = 0, \ \ \ \
\hat{\cal W}_{[2]}^0(h){\cal S}_{[3]}^\pm(h) =  \pm {3\sqrt{1+h^2}} \, {\cal
S}_{[3]}^\pm(h),
\nn \\
\hat{\cal W}_{[2]}^0(h) {\cal S}_{[21]}^0(h) = 0, \ \ \ \
\hat{\cal W}_{[2]}^0(h) {\cal S}_{[21]}^\pm(h) = \mp\frac{3\sqrt{1+h^2}}{2}\,{\cal
S}_{[21]}^\pm(h)
\ee
$h$-deformations of ${\cal W}_{[2]}^\bot$ and ${\cal W}_{[3]}$ are also easy to deduce.

\section{Conclusion}

In this paper we elaborated a little further on the suggestion of \cite{3Schur}
to define the analogue of Schur functions for plane partitions (nicknamed 3-Schurs)
through the recursion procedure.
Concretely, we provided detailed description of cut-and-join operators,
for which the 3-Schurs are the common eigenfunctions,
and used them to better explain
the reduction procedures, converting 3-Schur to the ordinary 2-Schur and
Macdonald polynomials, which depend only on the ordinary partitions (Young diagrams).
Most formulas are worked out for the first non-trivial case of "level two",
when there are three plane and two ordinary partitions, with two cubes and two boxes
respectively.
The Schur part of the story is explicitly lifted to level three,
but we do not discuss Macdonald part at this level --
on one hand it  rather straightforwardly follows the path, explained at level two,
on another hand it involves tedious calculations,
which would make more sense after the Schur part is independently checked and approved.
Moreover, at level three it can happen that 3-Schur functions are naturally describing
the entire multi-parametric set of Kerov functions \cite{MMkerov}, 
not only their Macdonal locus (at level two the difference is  not seen:
generic Kerov in this case is exactly Macdonald). 
We also did not reveal the relation to triple Macdonald polynomials of \cite{Z3pols},
which are directly applicable to network model studies \cite{network} --
still do not yet allow to formulate the $\left<character\right>\sim { character}$
relation \cite{MMcharreview}, what can actually require building a systematic
first-principle
theory of 3-Schur functions, which is the target of \cite{3Schur} and of the present paper.
This, however, can take time and quite some effort.
As to more local problems,
the main one at the moment is the lack of clear view on {\it triality}
between $q,t^{-1},tq^{-1}$, which is expected to naturally occur
if reduction from 3-Schur to Macdonalds is properly described --
but so far it does not.
All these open questions point to the obvious directions for further research
in this promising field.

\section*{Acknowledgements}

My work is partly supported by the grant of the Foundation for the Advancement of Theoretical
Physics
"BASIS", by RFBR grant  16-02-01021  and by the joint
grants  18-51-05015-Arm, 18-51-45010-Ind, 17-51-50051-YaF.


\begin{thebibliography}{12}


\bibitem{UFN3}
A. Morozov,   Sov. Phys. Usp. 35 (1992) 671;  Sov. Phys. Usp.  37 (1994) 1, hep-th/9303139 \!\!;
hep-th/9502091\!\!; hep-th/0502010 \\
A. Mironov, Int.J.Mod.Phys. A9 (1994) 4355 \,\,; Phys.Part.Nucl. 33 (2002) 537 \!\!; hep-th/9409190

\bibitem{GKMMM} 
A. Gorsky, I. Krichever, A. Marshakov, A. Mironov, A. Morozov,
Phys.Lett. B355 (1995) 466-474,  arXiv:hep-th/9505035  

\bibitem{CFT} A. Belavin, A. Polyakov, A. Zamolodchikov, Nucl.Phys. B241 (1984) 333-380\\
A. Zamolodchikov, Al. Zamolodchikov,
{\it Conformal field theory and critical phenomena in 2d systems}, 2009 \\
L. Alvarez-Gaume, Helvetica Physica Acta, 64 (1991) 361 \\
P. Di Francesco, P. Mathieu, D. Senechal, {\it Conformal Field Theory}, Springer, 1996 \\
A. Marshakov, A. Mironov and A. Morozov, Theor. Math. Phys. 164 (2010) 831, arXiv:0907.3946 \\
A. Mironov, S. Mironov, A. Morozov and An. Morozov, arXiv:0908.2064

\bibitem{Nekfun}
N. Nekrasov, Adv. Theor. Math. Phys. 7 (2004) 831–864 \\
R. Flume and R. Pogossian, Int. J. Mod. Phys. A18 (2003) 2541 \\
N. Nekrasov and A. Okounkov, hep-th/0306238 \\
N. Nekrasov and S. Shadchin, Commun.Math.Phys. 252 (2004) 359-391,   arXiv:hep-th/0404225 \\
A. Mironov and A. Morozov, Phys. Lett. B 680 (2009) 188, arXiv:0908.2190


\bibitem{AGT}
L. Alday, D. Gaiotto and Y. Tachikawa, Lett. Math. Phys. 91 (2010) 167–197, arXiv:0906.3219 \\
N. Wyllard, JHEP 0911 (2009) 002, arXiv:0907.2189 \\
A. Mironov and A. Morozov, Nucl. Phys. B825 (2009) 1–37, arXiv:0908.2569

\bibitem{confmamo} Vl. Dotsenko and V. Fateev, Nucl.Phys., B240 (1984) 312-348 \\
A. Marshakov, A. Mironov and A. Morozov, Phys. Lett. B 265 (1991) 99 \\
S. Kharchev, A. Marshakov, A. Mironov, A. Morozov and S. Pakuliak, Nucl. Phys. B 404 (1993)
717, hep-th/9208044 \\
R. Dijkgraaf and C. Vafa, arXiv:0909.2453 \\
H. Itoyama, K. Maruyoshi and T. Oota, Prog. Theor. Phys. 123 (2010) 957, arXiv:0911.4244 \\
A. Mironov, A. Morozov, Sh. Shakirov, JHEP 1002 (2010) 030,  arXiv:0911.5721 \!\!;
Int.J.Mod.Phys. A25 (2010) 3173-3207, arXiv:1001.0563 \!\!;
J.Phys.A44:085401 (2011), arXiv:1010.1734 \!\!;
Int.J.Mod.Phys. A27 (2012) 1230001,  arXiv:1010.1734 \\
H. Itoyama and T. Oota, Nucl. Phys. B838 (2010) 298-330, arXiv:1003.2929 \\
A. Mironov, A. Morozov, Sh. Shakirov, A. Smirnov, Nucl.Phys. B855 (2012) 128-151,
 arXiv:1105.0948 \\
A. Morozov and  Y. Zenkevich, JHEP 1602 (2016) 098,  arXiv:1510.01896 \\
A. Mironov and A. Morozov, Phys.Lett. B773 (2017) 34-46,  arXiv:1707.02443

\bibitem{Macdonald}  I.G. Macdonald,
{\it Symmetric functions and Hall polynomials}, Oxford Science Publications, 1995

\bibitem{DIM}
J. Ding and K. Iohara, Lett. Math. Phys. 41 (1997) 181-193, q-alg/9608002 \\
K. Miki, J. Math. Phys. 48 (2007) 123520 \\
B. Feigin and A. Tsymbaliuk, Kyoto J. Math. 51 (2011) 831-854, arXiv:0904.1679 \\
B. Feigin, E. Feigin, M. Jimbo, T. Miwa and E. Mukhin, Kyoto J. Math. 51 (2011) 337-364,
arXiv:1002.3100 \\
B. Feigin, K. Hashizume, A. Hoshino, J. Shiraishi and S. Yanagida,
J. Math. Phys. 50 (2009) 095215, arXiv:0904.2291 \\
B. Feigin, E. Feigin, M. Jimbo, T. Miwa and E. Mukhin, Kyoto J. Math. 51 (2011) 365-392,
arXiv:1002.3113 \\
H. Awata, B. Feigin, A. Hoshino, M. Kanai, J. Shiraishi and S. Yanagida, arXiv:1106.4088 \\
B. Feigin, M. Jimbo, T. Miwa and E. Mukhin, Kyoto J. Math. 52 (2012) 621-659,
arXiv:1110.5310 \!\!;
J.Alg. 380 (2013) 78-108, arXiv:1204.5378 \!\!;
Adv.Math. 300 (2016) 229, arXiv:1309.2147 \!\!;  arXiv:1502.07194; arXiv:1603.02765 \\
H. Awata, B. Feigin and J. Shiraishi, arXiv:1112.6074 \\
H. Awata, H. Kanno, T. Matsumoto, A. Mironov, A. Morozov, An. Morozov, Y. Ohkubo, Y.
Zenkevich,
JHEP 07 (2016) 1-67,  arXiv:1604.08366 \!\!; Phys. Rev. D 96 (2017) 026021,  arXiv:1703.06084 \\
H. Awata, H. Kanno, A. Mironov, A. Morozov, K. Suetake, Y. Zenkevich,
JHEP 2018 (2018) 192,  arXiv:1712.08016

\bibitem{network}
A. Iqbal, N. Nekrasov, A. Okounkov and C. Vafa, JHEP 0804 (2008) 011, hep-th/0312022 \\
J.-E. Bourgine, Y. Matsuo and H. Zhang, arXiv:1512.02492 \\
T. Kimura and V. Pestun, arXiv:1512.08533 \!\!; arXiv:1608.04651 \!\!;  arXiv:1705.04410 \\
A. Mironov, A. Morozov, Y. Zenkevich, JHEP  05 (2016) 1-44, arXiv:1603.00304 \!\!;
Phys.Lett. B762 (2016) 196-208,  arXiv:1603.05467   \\
H. Awata, H. Kanno, A. Mironov, A. Morozov, An. Morozov, Y. Ohkubo and Y. Zenkevich,
JHEP 1610 (2016) 1-49,  arXiv:1608.05351 \!\!;
Nucl.Phys. B918 (2017) 358-385,  arXiv:1611.07304 \\
J. E. Bourgine, M. Fukuda, Y. Matsuo and R. D. Zhu, JHEP 1712 (2017) 015, arXiv:1709.01954 \\
F. Nieri, Y. Pan and M. Zabzine,  arXiv:1711.06150 \!\!; arXiv:1807.11900 \\
O. Foda and M. Manabe, arXiv:1801.04943

\bibitem{3Schur} A.Morozov, Phys.Lett. B785 (2018) 175-183,  arXiv:1808.01059 v3

\bibitem{genpols}
H. Awata, B. Feigin, A. Hoshino, M. Kanai, J. Shiraishi, S. Yanagida, arXiv:1106.4088 \\
A.~Morozov, A.~Smirnov,
   Lett.Math.Phys.\ {\bf 104} (2014) 585, arXiv:1307.2576 \\
S.~Mironov, An.~Morozov, Y.~Zenkevich,
JETP Lett.\  {\bf 99} (2014) 109, arXiv:1312.5732 \\
Y.~Ohkubo, arXiv:1404.5401 \\
B.~Feigin, M.~Jimbo, T.~Miwa, E.~Mukhin, arXiv:1502.07194 \\
Y.~Kononov and A.~Morozov,  Eur.Phys.J. C76 (2016) no.8, 424,  arXiv:1607.00615     \\
Y.~Zenkevich,  arXiv:1612.09570

 

\bibitem{Z3pols}
Y. Zenkevich,  arXiv:1712.10300

\bibitem{Nekmag} N.Nekrasov  arXiv:1712.08128 \\
N.Nekrasov and N.Piazzalunga,  arXiv:1808.05206

 


\bibitem{MMN1}  A. Mironov, A. Morozov, S. Natanzon,
Theor.Math.Phys. 166 (2011) 1-22, arXiv:0904.4227 \!\!; J.Geometry and Physics 62 (2012) 148-155,
arXiv:1012.0433  


\bibitem{MMkerov}
 S.V. Kerov, Func.An.and Apps. 25 (1991) 78-81 \\
A. Mironov and A. Morozov,  arXiv:1811.01184

 




\bibitem{tenmod}
J.B. Geloun, R. Gurau, V. Rivasseau, Europhys.Lett. 92 (2010) 60008, arXiv:1008.0354 \\
R. Gurau, V. Rivasseau, Europhys.Lett. 95 (2011) 50004, arXiv:1101.4182 \\
V. Bonzom, R. Gurau, A. Riello, V. Rivasseau, Nucl.Phys. B853 (2011) 174-195,
arXiv:1105.3122 \\
V. Bonzom, R. Gurau, V. Rivasseau, Phys.Rev. D85 (2012) 084037, arXiv:1202.3637 \\
R. Gurau, Nucl.Phys. B852 (2011) 592, arXiv:1105.6072;  arXiv:1203.4965 \!\!;
Nucl. Phys. B916 (2017) 386, arXiv:1611.04032; arXiv:1702.04228 \\
V. Bonzom, arXiv:1208.6216; JHEP 06 (2013) 062, arXiv:1211.1657 \\
E. Witten, arXiv:1610.09758 \\
I. Klebanov, G. Tarnopolsky, Phys.Rev. D 95 (2017) 046004, arXiv:1611.08915 \\
S. Carrozza, A. Tanasa, Letters in Mathematical Physics, 106(11) (2016) 1531-1559,
arXiv:1512.06718
\\
A. Jevicki, K. Suzuki, J. Yoon, JHEP, 07 (2016) 007, arXiv:1603.06246 \\
H. Itoyama, A. Mironov, A. Morozov, Phys.Lett. B771 (2017) 180-188, arXiv:1703.04983 \\
S. Das, A. Jevicki, K. Suzuki, arXiv:1704.07208 \\
K. Bulycheva, I. Klebanov, A. Milekhin, G. Tarnopolsky, arXiv:1707.09347 \\
P. Diaz and J. A. Rosabal, arXiv:1809.10153 [hep-th] \!\!;  arXiv:1810.02520 [hep-th]


\bibitem{NLA}
 I. Gelfand, M. Kapranov and A. Zelevinsky,
{\it Discriminants, Resultants and Multidimensional Determinants}, Birkhauser, 1994 \\
V. Dolotin and A. Morozov,
{\it Introduction to Non-Linear Algebra}, WS, Singapore 2007,  hep-th/0609022 \\
A.Morozov, Sh.Shakirov, arXiv:0911.5278




\bibitem{MMcharreview}
A. Mironov and A. Morozov, JHEP 1808 (2018) 163,  arXiv:1807.02409 \\
A. Morozov, A. Popolitov and Sh. Shakirov, Phys.Lett. B784 (2018) 342-344,
arXiv:1803.11401 \\
H. Itoyama, A. Mironov and A. Morozov,  arXiv:1808.07783 \\
see also \cite{MAMOchars2} for important earlier papers

\bibitem{MAMOchars2}
R. de Mello Koch, S. Ramgoolam, arXiv:1002.1634 \\
J. Ben Geloun, S. Ramgoolam, arXiv:1307.6490 \\
H. Itoyama, A. Mironov, A. Morozov,  JHEP, 06 (2017) 115, arXiv:1704.08648 \!\!;
Nucl.Phys. B932 (2018) 52-118, arXiv:1710.10027 \\
P. Diaz and S.J. Rey, JHEP 1802 (2018) 089,
arXiv:1706.02667 [hep-th] \!\!;
Nucl.Phys. B932 (2018) 254-277,
arXiv:1801.10506 [hep-th] \\
A. Mironov and A. Morozov, Phys.Lett. B771 (2017) 503-507, arXiv:1705.00976 \!\!;
Phys.Lett. B774 (2017) 210-216, arXiv:1706.03667 \\
R. de Mello Koch, D. Gossman, L. Tribelhorn, JHEP, 2017 (2017) 011, arXiv:1707.01455 \\
J. Ben Geloun, S. Ramgoolam, arXiv:1708.03524 \\
P. Diaz,    JHEP 1806 (2018) 140,
arXiv:1803.04471 [hep-th]

\end{thebibliography}
\end{document}